\documentclass[cntp]{ipart}
\usepackage{hyperref}
\arxiv{2302.01251}

\pdfoutput=1
\usepackage{bbm}
\usepackage{tikz}
\usetikzlibrary{calc}
\usepackage{graphicx,subcaption}
\usepackage{mathtools}
\usepackage{bm}
\usepackage{enumitem}
\usepackage{verbatim}
\usepackage[normalem]{ulem}

\newcommand{\fmt}{\frac{m}{2}}
\renewcommand{\fmt}{\lambda}

\newcommand{\bfmtb}{\bfmtb}
\renewcommand{\bfmtb}{\lambda}

\renewcommand{\Im}{\operatorname{Im}}

\DeclareMathAlphabet{\mathbfsf}{OT1}{cmss}{bx}{n}

\newcommand{\CF}{\mathcal{F}}

\newcommand{\CL}{\mathcal{L}}

\newcommand{\CO}{\mathcal{O}}

\newcommand{\CH}{\mathcal{H}}

\newcommand{\IR}{\mathbb{R}}
\newcommand{\IC}{\mathbb{C}}
\newcommand{\IZ}{\mathbb{Z}}
\newcommand{\IH}{\mathbb{H}}

\newcommand{\Zint}{\mathbb{Z}}

\newcommand{\Z}{\mathbb{Z}}
\newcommand{\C}{\mathbb{C}}
\newcommand{\R}{\mathbb{R}}

\renewcommand{\=}{\;= \;}

\newcommand\be{\begin{equation}}
\newcommand\ee{\end{equation}}
\newcommand\beq{\begin{equation}}
\newcommand\eeq{\end{equation}}
\newcommand\bea{\begin{eqnarray}}
\newcommand\eea{\end{eqnarray}}

\renewcommand{\a}{\alpha}
\renewcommand{\b}{\beta}

\renewcommand{\t}{\tau}

\newcommand{\wh}{\widehat}

\newcommand{\p}{\partial}

\renewcommand{\i}{{\rm i}}

\newcommand{\half}{\frac12}

\newcommand{\rme}{{\rm e}}

\newcommand{\dd}{{\rm d}}

\newcommand{\zbar}{\overline{z}}

\newcommand{\ii}{{\rm i}}

\renewcommand{\hat}{\widehat}
\renewcommand{\bar}{\overline}

\newcommand{\slz}{\text{SL}(2,\mathbb{Z})} 
\newcommand{\taubar}{\overline{\tau}} 

\newcommand{\vev}[1]{\bigl\langle {#1}  \bigr\rangle } 
\newcommand{\lb}{\left(}
\newcommand{\rb}{\right)}

\begin{document}

\begin{frontmatter}

\title{Modular symmetry of massive free fermions}

\begin{aug}

  \author{\fnms{Max} \snm{Downing}\ead[label=e1]{max.downing@kcl.ac.uk}},
  \address{Department of Mathematics, King's College London,
  The Strand, London WC2R 2LS,\\
  U.K.\\
\printead{e1}}
  
\author{\fnms{Sameer} \snm{Murthy}\ead[label=e2]{sameer.murthy@kcl.ac.uk}},
\address{Department of Mathematics, King's College London,
  The Strand, London WC2R 2LS,\\
  U.K.\\
\printead{e2}}

\and

\author{\fnms{Gerard} \snm{Watts}\ead[label=e3]{gerard.watts@kcl.ac.uk}}
\address{Department of Mathematics, King's College London,
  The Strand, London WC2R 2LS,\\
  U.K.\\
\printead{e3}}

\end{aug}

\begin{abstract}
 We construct an infinite set of conserved tensor currents of rank~$2n$, $n=1,2,\dots$, in the two-dimensional theory of 
free massive fermions, which are bilinear in the fermionic fields. 
The one-point functions of these currents on the torus depend on the modular parameter~$\t$ and spin structure~$(\a,\b)$.
 We show that, upon scaling the mass~$m$ so as to keep the combination~$m^2 \Im(\t)$ invariant, the one-point functions 
are non-holomorphic Jacobi forms of weights~$(2n,0)$ or~$(0,2n)$ and index~0, with respect to the modular 
parameter~$\t$ and elliptic parameter~$z=\a\t+\b$. 
In particular, we express the one-point functions as Kronecker-Eisenstein-type sums over the 
lattice~$\IZ\t+\IZ$, which makes the modular symmetry manifest.
We show that there is an action of three differential 
operators on these Jacobi forms which form an~$\mathfrak{sl}_2(\IR)$ Lie algebra.
Further we show that these Jacobi forms obey three differential
equations arising from the representation theory of the Jacobi group.
\end{abstract}


\end{frontmatter}


\section{Introduction}

The relation between two-dimensional 
conformal field theory (CFT) and modular forms is well-known, for example characters of a CFT are vector-valued modular forms 
and the partition function on a torus is invariant under modular transformations~\cite{CARDY1986186}. 
The rigorous mathematical development of this relation involves Vertex Operator Algebras (VOAs), as in~\cite{Zhu1996}, but it is firmly based in the expected properties of physical theories arising in statistical mechanics or string theory.
For conformal fields, the modular properties of expectation values of fields (observables) also follow from the general mathematical theory \cite{Zhu1996}.
The known modular properties of massless free fermions fall into this class.

One physical realisation of CFT is in terms of statistical models defined on a torus and their calculation in 
terms of traces through a transfer matrix approach. In the continuum limit, 
this becomes a trace over a suitable space of states in a quantum field theory. The modular property arises from the 
equality of these two different ways to calculate the same physical quantity. 
There is an alternative point of view in physics given by the functional integral, which---although 
not rigorous---leads to a much more intuitive understanding of the modular properties of the free 
fermion or, more generally, of any CFT. 
Modular transformations are 
simply a combination of large diffeomorphisms of the torus and 
scale transformations. The former is a classical symmetry of any diffeomorphism-invariant theory 
and the latter is a symmetry of any conformal field theory. 
The combination, therefore, is a symmetry transformation of the underlying functional integral of the CFT, 
and any observable of such a theory should therefore transform in a covariant manner under it. For example, 
the partition function of massless free fermions 
on a torus leads to theta functions, one-point functions lead to Eisenstein series, and so on. 

It has long been recognised that partition functions of massive theories should also have modular properties. 
Indeed, this was shown for the partition function of the off-critical Ising model---which is equivalent to the 
theory of massive free fermions---by Itzykson and Saleur in 1987 \cite{Saleur1987}, see also \cite{KLASSEN1991635} and more recently \cite{Kostov_2022}. 
The question arose again in connection with string theory when it was realised that strings in a PP-wave background in the light-cone Green-Schwarz formalism were equivalent to 
free massive fields \cite{Metsaev_2002} and modular covariance of open strings \cite{Bergman_2003} and 
closed strings \cite{Takayanagi:2002pi} was shown. This in turn led to massive Maass forms, defined in~\cite{Berg:2019jhh} and further explored in \cite{Berg}.

There is, however, no similar general mathematical theory of modular properties of massive field theories (the Ising results come from direct calculation), but we can still use the physical intuition from the path-integral approach. 
The main difference is that scale transformations are explicitly broken. In general this will correspond to a renormalisation group flow in the space of massive field theories, but in the case of free field theories (such as massive fermions) this simply corresponds to a change in the mass parameter, and the above intuition can very easily be extended.

In this paper we show that there is an infinite class of observables in the massive free fermion theory 
which lead to modular forms. These modular forms take the form of Kronecker-Eisenstein series, which 
are non-holomorphic deformations of the Eisenstein
series occurring in free fermion CFTs, and reduce to them in the massless limit. 
In the rest of this introduction, we briefly summarise the arguments of modular symmetry for massive 
fermion theories, focusing on one-point functions. We review the functional integral explanation of 
modular symmetry in Appendix~\ref{sec:PImodform}.

\medskip

In physics, a one-point function is an expression
\begin{align}
    \vev{\hat{\cal O}}_{\cal M}^{\cal T}\;,
\end{align}
which is determined by a quantum field theory ${\cal T}$, an operator ${\cal O}$ in that theory and the manifold ${\cal M}$ on which the theory is defined. If one considers a coordinate transformation, each of ${\cal T}$, ${\cal O}$ and ${\cal M}$ can change leading to identities of the form
\begin{align}
    \vev{\hat{\cal O}}_{\cal M}^{\cal T}
    = 
    \vev{\hat{\cal O}'}_{\cal M'}^{\cal T'}\;.
\end{align}
By suitable choices of $\{{\cal T,O,M}\}$ this will lead to functions with well-defined modular properties.

In this paper we always take ${\cal M}$ to be a  two-dimensional torus defined as the quotient of the complex plane 
$\mathbb C/( \mathbb Z R\tau + \mathbb Z R)$. 
Here the complex number~$\tau$ is the complex modulus while the positive real number~$R$ is the K\"ahler modulus.
We will denote a one-point function on such a torus by $\vev{\cdots}_{\tau,R}$. 
This is long understood to lead to modular forms in the case that ${\cal T}$ is a conformal field theory.
If $\cal T$ is a conformal field theory and the coordinate transformation is a combined scale, rotation and translation, by definition one has ${\cal T}={\cal T'}$. Taking $\cal O$ to be a ``scaling field''$\phi$ with conformal dimensions $(h,\bar h)$ leads, through combining suitable translations, rotations and scalings, to the result
\begin{align}\label{eq:cftmt}
    &\vev{\phi}_{\hat\tau,1}^{\cal T}
    \= (c\tau+d)^h(c\bar\tau+d)^{\bar h} \vev{\phi}_{\tau,1}^{\cal T}
    \\
    &\wh\t \; \equiv \; \frac{a\tau+b}{c\tau+d} \,,  
\qquad{\Big({\small
\begin{matrix}a&b\\c&d\end{matrix}}\Big)\in \slz }\,,
\nonumber
\end{align}
i.e. the function
\begin{align}
    \tau \mapsto \vev{\phi}_{\tau,1}^{\cal T}\;,
\end{align}
is a non-holomorphic modular form of weights $(h,\bar h)$. 

In this paper we are interested in generalisations of this construction when ${\cal T}$ is no longer a conformal field theory but depends on a mass scale $m$. 
In particular, we 
consider the theory of two-dimensional, two component
fermion~$\psi$ of mass $m$  (with components $\Psi$ and $\bar\Psi$). As explained in Appendix~\ref{Quantising fermion}, 
this can either be formulated in Lorentzian signature as a Majorana fermion, or in Euclidean signature as a fermion obeying a Majorana-like relation and which is
formally described by the Lagrangian from \cite{Ghoshal:1993tm}
\be \label{LagFF}
\CL \= \Psi \bar\p \Psi - \bar\Psi\partial\bar\Psi + m \Psi \bar\Psi  \,.
\ee
Under a scaling transformation, the kinetic terms in~\eqref{LagFF} are still invariant, but the mass term is not. 
However, the scaling can be easily absorbed into the mass term as
\be
m \to |c\t+d| \, m  \,.
\ee
This has the result that one has to pay close attention to the dependence of the theory ${\cal T}$ on the size $R$ of the torus and instead of \eqref{eq:cftmt} we instead find relations of the form
\begin{align}
    \vev{\phi}^{{\cal T}(\hat m)}_{\hat\tau,1}
    \= F \vev{\phi}^{{\cal T}(m)}_{\tau,1} \,.
\end{align}
where~$F$ contains factors arising from the rotational and scaling properties of the field $\phi$.

We now summarise the main results presented in this paper. In Section~\ref{sec:massiveferm} we describe suitable fields in the free Majorana fermion model. These fields, defined in \eqref{Tchoice}, are the components of a conserved tensor field which is bilinear in the fermions. They are denoted by $T_{(2n,\pm 2n)}$ for $n=0,1,\dots$. 

In Section~\ref{ssecfermbil} we calculate the one point functions on the torus of these fields in the different spin structures (where we have set $R=1$). The one point functions \eqref{Tpmlatticesum} are Kronecker--Eisenstein lattice sums that we record here for convenience
\begin{align}\label{eq:Monepoint}
\vev{ T_{(2n,\pm 2n)}}_{\tau,m;\a,\b}{=}\lb\frac{m}{2}\rb^{2n} \sideset{}'\sum_{(r,\ell)\in\mathbb{Z}^2}\left(\frac{\bar\tau r-\ell}{\tau r-\ell}\right)^{\pm n}
K_{2n}(m|r\tau - \ell|)\, \rme^{2\pi i(\alpha \ell+\beta r)} \;,
\end{align}
where $\a,\b \in \{0,\frac12\}$ and the prime indicates we are excluding the origin from the lattice sum. These are the non-holomorphic modular forms we study in the rest of the paper.

In Section~\ref{sec:Diracsymm} we extend this calculation to related fields in the free massive Dirac fermion for twisted sectors. The components of the conserved tensor are now denoted $T^D_{(2n,\pm2n)}$ and are defined in \eqref{eq:Dtensor}. The one point functions \eqref{TpmlatticesumDir} of these fields take the same form as \eqref{eq:Monepoint} except now $\a,\b\in\R$.

Finally, in Section~\ref{ssecmoddxfm} we derive the modular transformation properties of the Dirac fermion expectation values (equations \eqref{eq:Ttrans} and \eqref{eq:Strans})
\be
\begin{split}
    \vev{ T^D_{(2n,\pm 2n)}}_{\tau+1,m;\a,\b-\a}&\=\vev{ T^D_{(2n,\pm 2n)}}_{\tau,m;\a,\b}\;,\\
    \vev{ T^D_{(2n,2n)}}_{-1/\tau,|\t|m;-\b,\a} &\=\t^{2n}\vev{ T^D_{(2n,2n)}}_{\tau,m;\a,\b}\;,\\
    \vev{ T^D_{(2n,-2n)}}_{-1/\tau,|\t|m;-\b,\a}&\=\bar\t^{2n}\vev{ T^D_{(2n,-2n)}}_{\tau,m;\a,\b}\;.
\end{split}
\ee
The mass $m$ transforms as a weight $(1,1)$ modular form so we introduce an invariant mass $\mu \= m^2 \Im(\t) \= m^2\t_2$ and write our one point functions in terms of this new parameter.
The massless limit of our formulae is discussed in Section~\ref{sec:massless}.

In Section~\ref{sec:diffeq} we turn to the differential equations satisfied by the functions we found in Section~\ref{sec:massiveferm}. We show that the one point functions are massive Maass and massive Jacobi forms. (The relevant definitions were first given in \cite{Berg:2019jhh} and are repeated in Sections \ref{ssec:MF} and \ref{ssec:Jg} respectively.) In Section~\ref{ssec:lowerraise} we discuss an $\mathfrak{sl}_2(\R)$ action on the functions. The differential operators \eqref{eq:raiselower}
\be
    X_+ \= X_+^{(k)}\=(\t-\bar \t)\partial_\t + k\,, \qquad 
    X_- \= X_-^{(\bar k)}\=-(\t-\bar\t)\partial_{\bar\t} + \bar k\,,
\ee
act on modular forms of weight $(k,\bar k)$, and together with $Z = (k-\bar k)$ form a representation of $\mathfrak{sl}_2(\R)$ (see \eqref{eq:sl2Rrep}). They can then be used to move between the different one point functions \eqref{eq:Traiselower} 
\be
\begin{split}
&X_\pm\vev{ T_{(2n,\pm2n)}}_{\!\t,\mu;\a,\b} {=} -4\t_2\,\partial_\mu\vev{ T_{(2n+2,\pm(2n+2))}}_{\!\t,\mu;\a,\b}\;,\\
    &X_\mp\vev{ T_{(2n,\pm2n)}}_{\!\t,\mu;\a,\b} {=} \frac{1}{4\t_2}(-\mu^2\partial_\mu {+} 2\mu(n-1))\vev{ T_{(2n-2,\pm(2n-2))}}_{\!\t,\mu;\a,\b}\;.
\end{split}
\ee
In Section~\ref{ssec:MF} we give the relation to massive Maass forms. We define the Laplace operator \eqref{DelXXrel}
\be
    \Delta_{\tau}\=-\frac{1}{2}(X_+X_-+X_-X_+)\;,
\ee
and show 
\be
    \Delta_\t \vev{T_{2n,\pm 2n}}_{\t,\mu,\a,\b} = \left(-\mu^2 \partial_\mu^2 + 2\mu(n-1)\partial_\mu + n \right) \vev{T_{2n,\pm 2n}}_{\t,\mu,\a,\b} \;.
\ee
Hence the one point functions are massive Maass forms.

In Section~\ref{ssec:Jg} we discuss the differential equations arising from the Jacobi group. We combine the spin structure parameters $\a$ and $\b$ into a complex Jacobi variable $z = \a\t+\b$. Introducing the differential operators \eqref{eq:Ydiffop}
\be
    Y_{+} \= i\sqrt{\frac{\t-\bar\t}{2\ii}}\, \partial_z\;,\quad Y_- \= -i\sqrt{\frac{\t-\bar\t}{2\ii}}\, \partial_{\bar z} \,,
\ee
and modifying the $X_\pm$ operators \eqref{eq:Xdiffop}
\be
    X_+ = (\t-\bar\t)\partial_\t + (z-\bar z) \partial_z + k\,, \quad 
    X_- = -(\t-\bar\t)\partial_{\bar\t} - (z-\bar z) \partial_{\bar z} + \bar k\,,
\ee
gives us a representation of the Jacobi group. The Laplacian in the Jacobi variable \eqref{eq:z Laplace} is
\be
    \Delta_z = Y_+Y_- + Y_-Y_+ \;,
\ee
and the Casimir of the Jacobi group is
\be
    C_{\t,z} =\! \frac{1}{2}(Y_+^2X_- +\! X_-Y_+^2 -\! Y_-^2X_+ -\! X_+Y_-^2 + \!(k {-} \bar k)\!(Y_+Y_- +\! Y_-Y_+ \!)\!).
\ee
The one point functions are shown to satisfy
\begin{align}
    &C_{\t,z} \vev{T_{2n,\pm 2n}}_{\t,\mu,z} = 0\;,\\
    &\Delta_z \vev{T_{2n,\pm 2n}}_{\t,\mu,z} = -8\pi^2 \left(\mu \partial_\mu^2 - (2n-1)\partial_\mu \right) \vev{T_{2n,\pm 2n}}_{\t,\mu,z}\;,
\end{align}
and therefore are massive Jacobi forms.

Section~\ref{sec:conc} contains our conclusions. We include a series of appendices containing technical details.

\section{Massive free fermions \label{sec:massiveferm}}

We consider the theory of a massive two-dimensional 
Majorana fermion $\bigl(\bar\Psi(z,\zbar), \Psi(z,\zbar) \bigr)$ on the torus (see Appendix \ref{Quantising fermion} for what is meant by Majorana).
We use complex coordinates~$(z,\zbar)$ on the torus and use the 
notation~$\p = \p_z$, $\bar\p = \p_{\zbar}$. 
The  Lagrangian density is formally given by\footnote{This is the same as in \cite{Ghoshal:1993tm} but differs by a factor of 2 and the sign of $m$ from that in \cite{Saleur1987} so that here $m>0$ corresponds to the low-temperature phase.}
\begin{align}
\label{eq:massiveLag}
 \CL \=   
 \Psi\bar\partial\Psi - \bar\Psi \,\partial\bar\Psi + m\Psi\bar\Psi\;.
\end{align}
In this theory, consider a tensor field which is bilinear in the fermions and 
has fixed charge under scaling and rotation. 
As we show in Appendix~\ref{Current Uniqueness}, 
the components of such a tensor field are uniquely fixed to be, for $n = 1,2,\dots$,
\be \label{Tchoice}
T_{(2n,-2n)} =\frac{i}{2}\bar\Psi\,\bar\partial^{2n-1}\bar\Psi\,, \;\;
T_{(2n,2n)} =-\frac{i}{2}\Psi\partial^{2n-1}\Psi\,,\;\;
T_{(2,0)} = \frac{i\lambda}{2} \Psi\bar\Psi \,,
\ee
up to total derivatives and powers of the mass $m$ (where $\fmt=m/2$). 
Here the subscripts label the scaling and rotation eigenvalues, respectively, of the tensor fields.
Before evaluating the one-point function of 
these tensor fields, we make some comments about why we discuss these tensor fields in particular. 

\begin{figure}[htb]
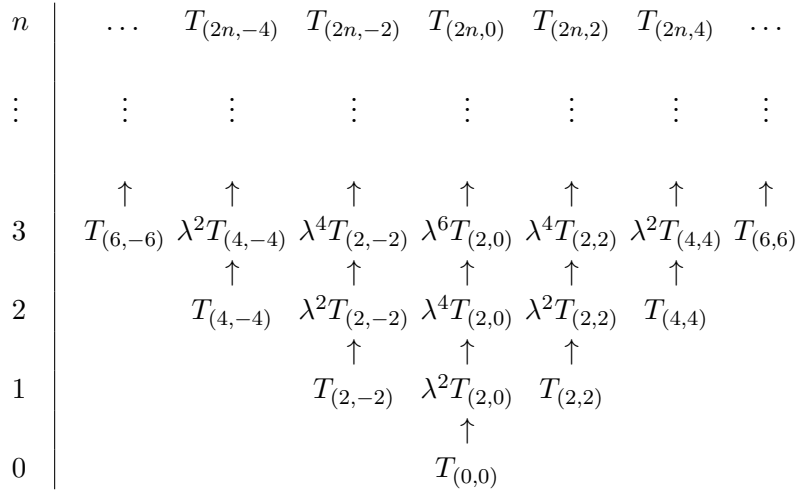

\[
{\renewcommand{\arraystretch}{1}
\begin{array}{p{0.5cm}|@{~~}c@{~}ccccccc}
$n$&&\dots&T_{(2n,-4)}&T_{(2n,-2)}&T_{(2n,0)}&T_{(2n,2)}&T_{(2n,4)}&\dots\\
&&&&&&&&\\
\vdots&&\vdots&\vdots&\vdots&\vdots&\vdots&\vdots&\vdots\\
&&\uparrow&\uparrow&\uparrow&\uparrow&\uparrow&\uparrow&\uparrow\\
3&&T_{(6,-6)}&\bfmtb^2T_{(4,-4)}&\bfmtb^4T_{(2,-2)}&\bfmtb^6 T_{(2,0)}&\bfmtb^4T_{(2,2)}&\bfmtb^2T_{(4,4)}&T_{(6,6)}\\
&&&\uparrow&\uparrow&\uparrow&\uparrow&\uparrow&\\
2&&&T_{(4,-4)}&\bfmtb^2T_{(2,-2)}&\bfmtb^4 T_{(2,0)}&\bfmtb^2T_{(2,2)}&T_{(4,4)}&\\
&&&&\uparrow&\uparrow&\uparrow&&\\
1&&&&T_{(2,-2)}&\bfmtb^2 T_{(2,0)}&T_{(2,2)}&&\\
&&&&&\uparrow&&&\\
0&&&&&T_{(0,0)}&&&
\end{array}}
\]
\caption{The fields $T_{(2n,k)}$ ($n=1,2,\dots$, $k=-2n,-2n+2,\dots ,2n$) are components of a conserved tensor field of rank~$2n$. 
At each level there are two new components, $T_{(2n,\pm 2n)}$, and 
the other components are related to tensors of lower rank with 
additional factors of the mass, $m$, so as to have the correct dimension (we use $m=2\lambda$ to simplify the expressions). 
The field~$T_{(0,0)}$ is the logarithmic derivative of the partition function and naturally sits at the apex of this cone. 
}\label{fig:tensor pyramid}
\end{figure}

Firstly, one can check (see Appendix~\ref{Current Uniqueness} for details)
that the fully symmetric tensor with the following components, up to total derivatives, for $n=1,2,\dots$,
\be
T_{(2n,k)}\= \begin{cases} \frac{i}{2}\bfmtb^{2n+k}\bar\Psi\,{\bar\partial}^{-k-1}\bar\Psi\;,\qquad & k\=-2n,-2n+2,\dots,-2\;,\\
\frac{i}{2}\bfmtb^{2n-1}\Psi\bar\Psi\;,\qquad & k\=0\;,\\
-\frac{i}{2}\bfmtb^{2n-k}\Psi\partial^{k-1}\Psi & k\=2,4,\dots,2n \,,
\end{cases}
\ee
defines a conserved current in the free fermion theory. 
Note that we have
\begin{align}
T_{(2n+2,k)} \= \bfmtb^2 T_{(2n,k)}\;,\;\;\;\;
k\=0,\dots,2n \,,
\label{eq:level relations}
\end{align}
and therefore at each level we only have two new components in the tensor,
\begin{align} \label{physdefTn}
T_{(2n,-2n)}\=\frac{i}{2}\bar\Psi\,{\bar \partial}^{2n-1}\bar\Psi\;,\;\;\;\;
T_{(2n,2n)}\=-\frac{i}{2}\Psi\partial^{2n-1}\Psi\,,
\end{align}
i.e., the first two fields in~\eqref{Tchoice}, 
and the other components come from the lower tensors with appropriate factors of the mass. For later convenience we also introduce the 1 component tensor
\be
    T_{(0,0)} \= \frac{i}{2\lambda} \Psi\bar\Psi\;.
\ee
All these fields form a cone structure as shown in Figure~\ref{fig:tensor pyramid}, with the edges consisting of~$T_{(2n,\pm 2n)}$.
Note that the fields~$T_{(2n,\pm 2n)}$ are not holomorphic
(or anti-holomorphic) functions,  i.e., they depend on~$z$ and~$\zbar$.
One could also choose tensors which are not completely symmetric. 
For example, the anti-symmetric tensor with two components is completely determined by the vector (spin 1).
Finally, note that in the massless limit of the theory the only components of this tensor that survive are the first two tensors of~\eqref{Tchoice}.

In \cite{Berg:2019jhh} the partition function and its relation to massive Maass forms and massive Maass-Jacobi forms is studied in detail. The one point function of the middle component, $T_{(0,0)}$, is 
the logarithmic derivative with respect to mass of the partition function (see~\eqref{eq:dmlogZ}),  
as expected from the functional integral. Further, the one point functions of $T_{(2,\pm 2)}$ 
are the logarithmic derivatives of the partition function (see \eqref{eq:XlogZ}) with respect to the differential operators $X_\pm$ (in~$\t, \taubar$) given in~\eqref{eq:raiselower}. 

\subsection{One-point functions of fermion bilinears}
\label{ssecfermbil}

Now we calculate the one-point functions of~$T_{(2n,\pm 2n)}$ and $T_{(0,0)}$ in the Hamiltonian formalism. 
We begin by quantising the free fermion. 
Consider the 2 component fermionic field~$(\bar\Psi(x), \Psi(x))$
obeying the general twisted boundary condition
\begin{align}
\label{eq:fftwistbc}
    \Psi(z+R,\bar z+R) = \rme^{2\pi i\alpha} \, \Psi(z,\bar z)\;, \; \;
    \bar\Psi(z+R,\bar z+R) = \rme^{2\pi i\alpha} \, \bar\Psi(z,\bar z)\;,
\end{align}
for $\alpha=0$ or $\tfrac{1}{2}$. Such a field has the following Fourier expansion, 
\be
\Psi(z,\bar z) \= \sum_{k\in\mathbb{Z}+\alpha}
{\cal A}_k(z-\bar z)\, \rme^{\frac{\pi i k}{R}(z+\bar z)}\;,
\ee
where $\mathcal{A}_k$ is a 2 component spinor that only depends on $z-\bar z$ (similarly for $\bar\Psi$). ${\cal A}_k$ is then fixed by the equations of motion. The solution to the equations of motion following from the Lagrangian~\eqref{eq:massiveLag} 
is, upon imposing a Majorana-like condition explained in appendix \ref{Quantising fermion},
\be
\begin{split}
\label{psi12Fourier}
&\bar\Psi(z,\bar z) {=}\sqrt{\frac{2\pi}{R}} \!\!\sum_{k \in\mathbb{Z}+\alpha}\!\!\! \sqrt{\tfrac{\omega_k^-}{2\omega_k}} \!\left(\! \Psi_k e^{\frac{i\pi}{4}} e^{\frac{i}{2}(\omega_k^+z-\omega_k^-\bar z)} {+} \Psi_k^\dagger e^{-\frac{i\pi}{4}} e^{-\frac{i}{2}(\omega_k^+z-\omega_k^-\bar z)}\!\right)\,,\\
&\Psi(z,\bar z) {=}\sqrt{\frac{2\pi}{R}} \!\!\sum_{k \in\mathbb{Z}+\alpha}\!\!\! \sqrt{\tfrac{\omega_k^+}{2\omega_k}} \!\left(\! \Psi_k e^{-\frac{i\pi}{4}} e^{\frac{i}{2}(\omega_k^+z-\omega_k^-\bar z)} {+} \Psi_k^\dagger e^{\frac{i\pi}{4}} e^{-\frac{i}{2}(\omega_k^+z-\omega_k^-\bar z)}\!\right)\,,
\end{split}
\ee
where
\begin{equation}\label{eq:omegapm}
\omega_k\=\sqrt{m^2 + \left(\frac{2\pi k}{R}\right)^2}\;,\qquad \omega_k^\pm\=\omega_k \pm \frac{2\pi k}{R}\;.
\end{equation}
In the quantum theory one imposes the anticommutation relations
\begin{align}
\label{psipsibarac}
    \{\Psi_k,\Psi_\ell^\dagger\}
    \=\delta_{k,\ell} \,,
\end{align}
with all other anti commutators vanishing. In terms of the operators~$\Psi_k$, 
the Hamiltonian and momentum of the free fermion theory are given by
\be
\begin{split}\label{eq:HandP}
H &\=\sum_{k\in\mathbb{Z}+\alpha} \omega_k \left( \Psi_k^\dagger\Psi_k   - \frac{1}{2} \right)
\=\sum_{k\in\mathbb{Z}+\alpha} \omega_k \,\Psi_k^\dagger\Psi_k + E_0 \;, \\ 
P & \=\sum_{k\in\mathbb{Z}+\alpha} \frac{2\pi k}{R} \, \Psi_k^\dagger\Psi_k\;.
\end{split}
\ee
The zero-point energy is formally given by the series~$-\frac12 \sum_{k \in \IZ + \a} \omega_k$, which is, a priori, divergent.  
One can regularise this sum, and this leads to
\be\label{eq:E0}
E_0 \= \frac{m}{\pi}\sum_{\ell=1}^\infty \frac{1}{\ell}K_1(mR \ell) \cos(2\pi \ell\a) \,,
\ee
where~$K_1(x)$ is a modified Bessel function of the second kind. 
We present the calculation in Appendix~\ref{regularisation}. 
As it turns out, this value of~$E_0$ is precisely the one consistent with modular invariance.

\bigskip

We define the \emph{trace in the twisted sector~$\alpha, \beta$} for any operator~$\CO$ in the theory as
\be 
\text{Tr}_{\alpha,\beta}\; \CO 
\= \text{Tr}_{\CH_\alpha} \, \rme^{-2\pi i\bigl(\beta+\tfrac{1}{2}\bigr)F}\; \CO  \,,
\ee
where the Hilbert space~$\CH_\a$ is that of the free fermion with the twisted boundary conditions~\eqref{eq:fftwistbc} and the operator $\rme^{-2\pi i\bigl(\beta+\tfrac{1}{2}\bigr)F}$ obeys the following condition for all fermion modes~$\Psi_k$
\begin{align}\label{eq:betabc}
    \Psi_k\rme^{-2\pi i\bigl(\beta+\tfrac{1}{2}\bigr)F}=-\rme^{2\pi i\b}\rme^{-2\pi i\bigl(\beta+\tfrac{1}{2}\bigr)F}\Psi_k\;.
\end{align}
We define the partition function of the theory in the twisted sector~$\a,\b$ as
\be \label{defpartfnab}
Z_{\a,\b}(a,L,R) \= \text{Tr}_{\alpha,\b}\,
\exp \bigl( -LH + iaP \bigr)  \,.
\ee
Our goal is to calculate the normalised one-point functions 
\begin{align}\label{T1ptnorm}
&\vev{ T_{(2n,\pm 2n)}}_{a,L,R,m; \a,\b}
\= \frac{1}{Z_{\a,\b}}\;
\text{Tr}_{\alpha,\beta} \, \Bigl(  T_{(2n,\pm 2n)} \, 
\exp \bigl(-LH + iaP \bigr) \Bigr)\,,\\
&\vev{ T_{(0,0)}}_{a,L,R, m; \a,\b}
\= \frac{1}{Z_{\a,\b}}\;
\text{Tr}_{\alpha,\beta} \, \Bigl(  T_{(0,0)} \, 
\exp \bigl(-LH + iaP \bigr) \Bigr)\,.
\end{align}
Here the dependence on the length scales $a, L, R$ and the spin structure~$\a, \b$ is explicit on the right-hand side. 
The $m$-dependence of the right-hand side enters through that of the spectrum of the massive fermion theory.

\bigskip

We first gather some basic building blocks. 
A short calculation, using the anticommutation relations~\eqref{psipsibarac}, shows that
for any~$k,\ell \in \IZ+\a$ we have 
\be
\begin{split}
\label{eq:bilineartrace}
&\text{Tr}_{\alpha,\beta}(\Psi_k^\dagger\Psi_\ell \, \rme^{-LH + iaP)})
\= \delta_{k,\ell} \, \frac{-\rme^{-L\omega_k + 2\pi i\frac{a}{R}k+2\pi i\beta}}{1-\rme^{-L\omega_k + 2\pi i\frac{a}{R}k+2\pi i\beta}} \, Z_{\a,\b}(a,L,R) \,,\\
&\text{Tr}_{\alpha,\beta}(\Psi_k\Psi_\ell \, \rme^{-LH + iaP})\=0\,,
\qquad \text{Tr}_{\alpha,\beta}(\Psi_k^\dagger\Psi_\ell^\dagger \, \rme^{-LH + iaP}) \= 0 \,.
\end{split}
\ee
Next we expand the tensor field~\eqref{Tchoice} in terms of the 
operators~$\Psi_k$ using the Fourier expansion~\eqref{psi12Fourier}. Each term in this expansion is a bilinear in the operators~$\Psi_k$ and $\Psi^\dagger_k$. 
From~\eqref{eq:bilineartrace} it is clear that the trace of the bilinears of the type~$\Psi_k\Psi_\ell$,~$\Psi^\dagger_k\Psi^\dagger_\ell$ for any~$k,\ell \in \IZ+\a$, and~$\Psi^\dagger_k\Psi_\ell$, $k \neq \ell$ vanish. From the anticommutation relation~\eqref{psipsibarac}, we have that~$\Psi_k\Psi_\ell^\dagger$, $k \neq \ell$ also vanishes. 
Therefore, the terms that contribute to the trace are of the form~$\Psi_k^\dagger\Psi_k$  and~$\Psi_k\Psi_k^\dagger$,  $k \in \IZ+\a$. Again using the anticommutation relation~\eqref{psipsibarac}, we write the latter term 
as~$\Psi_k\Psi_k^\dagger=-\Psi_k^\dagger\Psi_k+1$. This last use of the anti commutation relations gives us a constant term that is formally a divergent series. One can give this constant a finite value either through regularisation of the divergent sum or by requiring the one point functions to be modular forms. Both approaches lead to the same value.

We summarise the above considerations in the following equations 
\begin{align}
    &
    \vev{ T_{(2n,\pm 2n)}}_{a,L,R, m; \a,\b}
\=\biggl{\langle} \frac{2\pi(-1)^{n+1}}{2^{2n}R}\sum_{k\in\mathbb{Z}+\alpha}\frac{(\omega_k^\pm)^{2n}}{\omega_k}\, \Psi_k^\dagger\Psi_k 
\biggr{\rangle}_{a,L,R, m; \a,\b}
+ C \,,\label{eq:T(2n,2n)}\\
    &
    \vev{ T_{(0,0)}}_{a,L,R, m; \a,\b}
\=\biggl{\langle}-\frac{2\pi}{R} \sum_{k\in\mathbb{Z}+\alpha}\frac{1}{\omega_k}\, \Psi_k^\dagger\Psi_k 
\biggr{\rangle}_{a,L,R, m; \a,\b} + C \,,\label{eq:T(0,0)}
\end{align}
where~$C$ is the constant that is formally given by the divergent series
\be
    \frac{2\pi(-1)^n}{2^{2n}R}\sum_{k\in \Z+\a} \frac{(\omega_k^\pm)^{2n}}{\omega_k}\;.
\ee
As shown in Appendix~\ref{regularisation}, this constant can be regularised to give
\be
    C\=\frac{m^{2n}}{2^{2n-1}}\sum_{\ell = 1}^\infty K_{2n}(mR \ell) \cos(2\pi \ell \a)\;,
\ee
which respects the modular properties of the one point function. We can set $n = 0$ in \eqref{eq:T(2n,2n)} to obtain \eqref{eq:T(0,0)} so we don't have to consider $\vev{T_{(0,0)}}_{\t,m,\a,\b}$ separately. 
Now we use \eqref{eq:bilineartrace} to calculate the trace of each term in this sum, in order to obtain
\begin{align}
    \vev{ T_{(2n,\pm2n)}}_{a,L,R, m; \a,\b} = \frac{2\pi (-1)^n}{2^{2n}R}  \sum_{k\in\mathbb{Z}+\alpha} \frac{(\omega_k^\pm)^{2n}}{\omega_k}\text{Li}_0\left(e^{-L\omega_k+2\pi i\frac{a}{R}k+2\pi i\beta}\right) +  C\,,
\end{align}
where~$ \text{Li}_0 (x)=x/(1-x)$ 
is the polylogarithm of order~$0$. We consider the theory on a torus with modular parameter $\t$, given by
\be
    \t \= \t_1 + i\t_2 \= \frac{a}{R} + i\frac{L}{R}\;.
\ee
Furthermore we fix the length of the spatial slice to be $R = 1$. The value of the one point functions on such a torus will now be a function of the modular parameter $\t$, the mass $m$ and the spin structure $(\a,\b)$, i.e.
\be\label{eq:onepoint}
    \vev{ T_{(2n,\pm2n)}}_{\t, m; \a,\b}\=\frac{2\pi (-1)^n}{2^{2n}}\sum_{k\in\mathbb{Z}+\alpha}\frac{(\omega_k^\pm)^{2n}}{\omega_k}\text{Li}_0\left(e^{-\t_2\omega_k + 2\pi i\t_1 k+2\pi i\beta}\right) + C\;,
\ee
It is clear from~\eqref{eq:onepoint} that the one-point functions are separately periodic in~$\a$ and~$\b$ with period~1,
and hence can be written as a double Fourier series.
We present the calculation of the Fourier coefficients in Appendix~\ref{Bessel sum}. 
The result for the special cases~$\a,\b \in \{0,\frac12\}$ is 
\begin{align} \label{Tpmlatticesum}
\vev{ T_{(2n,\pm 2n)}}_{\tau,m;\a,\b}\=\lb\frac{m}{2}\rb^{2n} \sideset{}'\sum_{(r,\ell)\in\mathbb{Z}^2}\left(\frac{\bar\tau r-\ell}{\tau r-\ell}\right)^{\pm n}
K_{2n}(m|r\tau - \ell|)\, \rme^{2\pi i(\alpha \ell+\beta r)} \;,
\end{align}
where the prime indicates we are excluding the origin from the lattice sum. We recognise this expression as a Kronecker--Eisenstein lattice sum, which also makes the
modular transformation property manifest. Indeed, the one-point 
functions $\vev{T_{(2n,\pm 2n)}}_{\t,m;\a,\b}$ for~$\a,\b \in \{0,\frac12\}$ are modular 
forms of weight~$(2n,0)$ and~$(0,2n)$, respectively.
We note\footnote{We thank M.~Berg for pointing this out.} that the same expressions appear in \cite{BergPersson}.

For generic~$\a$, $\b$, this is not true. As we see from \eqref{eq:halflattice}, in the general case one obtains a half-lattice sum because the coefficients with~$r<0$ vanish. 
This is related to the fact that we have a theory of Majorana fermions. In such a theory the boundary conditions must respect the reality condition on the fields. Hence we can only have $\alpha,\beta \in \{0,\tfrac{1}{2}\}$ in \eqref{eq:fftwistbc} and \eqref{eq:betabc}.
Indeed, this problem is solved by considering a theory of massive Dirac fermions, wherein one obtains full lattice sums for arbitrary spin structures. 
We present some details of the one-point functions in the Dirac theory,
followed by a general discussion of the modular properties in the following two subsections.

\subsection{Dirac fermions and symmetrisation in $\a,\b$}\label{sec:Diracsymm}

We now consider the two-component Dirac fermion~$(\Psi,\overline\Psi)$,  with the Lagrangian \eqref{eq:DiracLag}. 
The mode expansions for the two components are
\be\begin{split}
&\overline{\Psi}(z,\bar z) =
\sqrt{\frac{2\pi}{R}} \!\!\sum_{k \in\mathbb{Z}+\alpha}\!\!
\left(\!\! \sqrt{\tfrac{\omega_k^-}{2\omega_k}}\Psi_k e^{\frac{i\pi}{4}}e^{\frac{i}{2}(\omega_k^+z -\omega_k^-\bar z)}
+\sqrt{\tfrac{\omega_k^+}{2\omega_k}}\chi_k^\dagger e^{-\frac{i\pi}{4}}e^{\frac{i}{2}(\omega_k^+\bar z -\omega_k^-z)} \!\!\right)\,,\\
&\Psi(z,\bar z) =
\sqrt{\frac{2\pi}{R}} \!\!\sum_{k \in\mathbb{Z}+\alpha}\!\!
\left(\!\! \sqrt{\tfrac{\omega_k^+}{2\omega_k}}\Psi_k e^{-\frac{i\pi}{4}}e^{\frac{i}{2}(\omega_k^+z -\omega_k^-\bar z)}
+\sqrt{\tfrac{\omega_k^-}{2\omega_k}}\chi_k^\dagger e^{\frac{i\pi}{4}}e^{\frac{i}{2}(\omega_k^+\bar z -\omega_k^-z)} \!\!\right)\,,
\end{split}
\ee
with $\omega_n$ and $\omega_n^\pm$ as in~\eqref{eq:omegapm}. 
These fields have the periodicity condition \eqref{eq:fftwistbc}
but now for any~$\a$. As before we set $R = 1$, and after canonical quantisation the anti-commutators 
for the modes $\Psi_k$ and $\chi_k$ are the same as \eqref{psipsibarac}.
The Hamiltonian, $H$, and momentum, $P$, now take the form
\be
H = \sum_{k\in\mathbb{Z}+\a}\omega_k(\Psi_k^\dagger\Psi_k+\chi^\dagger_k\chi_k)+2E_0\;,\;\;
P = 2\pi \sum_{k\in\mathbb{Z}+\a} k(\Psi_k^\dagger\Psi_k-\chi^\dagger_k\chi_k)\;,
\ee
where $E_0$ is given in \eqref{eq:E0}. As above we define the trace in the twisted sector $\a,\b$ for any operator $\CO$ as
\be 
\text{Tr}_{\alpha,\beta}\; \CO 
\= \text{Tr}_{\CH_\alpha} \, \rme^{-2\pi i\left(\beta+\tfrac{1}{2}\right) F}\; \CO  \,,
\ee
where the charge~$F$ obeys
\begin{align} 
    \Psi_k \, \rme^{-2\pi i\bigl(\beta+\tfrac{1}{2}\bigr)F} \=-\rme^{2\pi i\b}\rme^{-2\pi i\bigl(\beta+\tfrac{1}{2}\bigr) F}\Psi_k\;,\quad 
\\    \chi_k \, \rme^{-2\pi i\bigl(\beta+\tfrac{1}{2}\bigr)F }\=-\rme^{-2\pi i\b}\rme^{-2\pi i\bigl(\beta+\tfrac{1}{2}\bigr)F }\chi_k\;.
\end{align}
Consider the bilinear currents
\begin{align}\label{eq:Dtensor}
    T_{(2n,2n)}^D \=-\frac{i}{2}\Psi^\dagger\partial^{2n-1}\Psi\;,\qquad T_{(2n,-2n)}^D\=\frac{i}{2}\bar\Psi^\dagger\bar\partial^{2n-1}\bar\Psi\;,
\end{align}
where the superscript $D$ denotes the fact these are now Dirac fermions instead of Majorana. As in \eqref{eq:bilineartrace} the only non-vanishing traces are over the bilinears $\Psi_k^\dagger\Psi_k$ and $\chi_k^\dagger\chi_k$, hence the expectation values on the torus with modular parameter $\t$ are
\be
\begin{split}\label{eq:ToneptD}
    &\vev{T^D_{(2n,\pm 2n)}}_{\!\tau, m; \a,\b} =\\
    &\frac{\pi (\!-1\!)^n}{2^{2n}} \!\!\!\!\sum_{k\in\mathbb{Z}+\alpha}\!\!\!\! \left(\!\! \frac{(\!\omega_k^\pm)^{2n}}{\omega_k}\text{Li}_0 \!\!\left(\! e^{-\tau_2\omega_k + 2\pi i\tau_1k + 2\pi i\beta} \!\right)\! {+} \frac{(\!\omega_k^\mp)^{2n}}{\omega_k}\text{Li}_0 \!\!\left(\! e^{-\tau_2\omega_k - 2\pi i\tau_1k - 2\pi i\beta} \!\right)\!\!\!\right)\! {+} C,
\end{split}
\ee
Upon expressing this equation as a Fourier expansion (see Appendix~\ref{Bessel sum}), 
and taking 
\be
    C\= \frac{m^{2n}}{2^{2n-1}}\sum_{\ell = 1}^\infty K_{2n}(m \ell) \cos(2\pi \ell \a)\;,
\ee
we obtain
\be\label{TpmlatticesumDir}
    \vev{T^D_{(2n,\pm 2n)}}_{\!\tau, m; \a,\b} {=} \lb\!\frac{m}{2}\! \rb^{2n} \!\!\!\sideset{}'\sum_{(r,\ell)\in\mathbb{Z}^2} \!\!\!\left(\frac{\bar\t r-\ell}{\t r-\ell}\right)^{\pm n}\! K_{2n}(m|\t r {-} \ell|) \, \rme^{2\pi i(\a \ell+\b r)}\;,
\ee
where again the origin is excluded from the sum.

Thus we obtain a full lattice sum expression for the one-point function for massive Dirac fermions for any~$\a,\b \in \IR$. 
Note that~\eqref{TpmlatticesumDir} 
can be expressed as a symmetrised sum of two one point functions of the Majorana theory, i.e.,
\be
 \vev{T^D_{(2n,\pm 2n)}}_{\tau,m;\a,\b} \=   
\frac{1}{2}\left(\vev{ T_{(2n,\pm 2n)}}_{\tau,m;\a,\b}+\vev{ T_{(2n,\pm 2n)}}_{\tau,m;-\a,-\b}\right) \,,
\ee
In the special case~$\a,\b \in \{0,\half\}$,
the two terms on the right-hand side of this equation are equal, which explains why we were 
able to write the corresponding expressions in the Majorana theory~\eqref{Tpmlatticesum} 
as full lattice-sums. 

\subsection{Modular transformation property}
\label{ssecmoddxfm}

We now discuss the modular properties of the one-point functions\\
$\vev{T^D_{(2n,\pm 2n)}}_{\tau,m;\a,\b}$. The expressions \eqref{TpmlatticesumDir} as Kronecker--Eisenstein 
lattice sums make the modular transformation property manifest. Indeed, the 
transformation~$\t \to \t +1$, $(\a,\b) \to (\a,\b-\a)$ is equivalent to the relabelling~$(r,\ell) \to (r, \ell -r)$
in the lattice sum, i.e.
\begin{align}\label{eq:Ttrans}
    \vev{ T^D_{(2n,\pm 2n)}}_{\tau+1,m;\a,\b-\a}\=\vev{ T^D_{(2n,\pm 2n)}}_{\tau,m;\a,\b}\;,
\end{align}
Similarly the transformation~$\t \to -1/\t$, $(\a,\b) \to (-\b,\a)$ and $m\to |\t|m$ is equivalent 
to the relabelling~$(r,\ell) \to (-\ell, r)$ in the lattice sum with an overall factor 
of~$\t^{2n}$ for~$T^D_{(2n,2n)}$ and ~$\taubar^{2n}$ for~$T^D_{(2n,-2n)}$, i.e.
\be\label{eq:Strans}
\begin{split}
    \vev{ T^D_{(2n,2n)}}_{-1/\tau,|\t|m;-\b,\a} &\=\t^{2n}\vev{ T^D_{(2n,2n)}}_{\tau,m;\a,\b}\;,\\
    \vev{ T^D_{(2n,-2n)}}_{-1/\tau,|\t|m;-\b,\a}&\=\bar\t^{2n}\vev{ T^D_{(2n,-2n)}}_{\tau,m;\a,\b}\;.
\end{split}
\ee
This shows that~$\vev{T^D_{(2n,\pm 2n)}}_{\t,m;\a,\b}$ are modular forms of weight $(2n,0)$ and $(0,2n)$, respectively.

Recall that for Majorana fermions we must restrict to $\a,\b\in\{0,\tfrac{1}{2}\}$. 
This is consistent because in this case the modular transformations transform the spin structures 
into ones within the same set.

Note that the real parameters~$\a$, $\b$ can be packaged into one complex parameter~$z=\a\t+\b$,
which is naturally interpreted as a background 
gauge field coupling to the fermion number current \cite{Kraus:2006wn}, 
such that the modular transformation acts on~$z$ as~$z \mapsto z/(c\t+d)$.
The one-point functions~$\vev{T^D_{(2n, 2n)}}$, $\vev{T^D_{(2n, -2n)}}$ as functions of~$(\t,z)$
transform as (non-holomorphic) Jacobi forms of weights~$(2n,0)$, $(0,2n)$, respectively.
It is clear that they are invariant under the elliptic 
transformations~$z \to z+ \lambda \t + \mu$, $\lambda,\mu \in \IZ$, i.e. they are index~0 Jacobi forms.

Finally, since under the $S$ transform $\t\to-1/\t$, we have $m\rightarrow|\t|m$ and 
under the $T$ transform $\t\to\t+1$, $m$ is invariant, $m^2$ is trivially a weight $(1,1)$ 
modular form. The rescaled mass parameter
\be\label{eq:invarmass}
\mu \= m^2 \Im(\t) \= m^2\t_2\;,
\ee
is therefore invariant under modular transformations, and we can therefore 
study the one point functions as functions of~$\tau, \bar\tau$ for fixed $\mu$.
In Appendix \ref{Current Uniqueness} we show that, up to total derivatives, all bilinear currents are of the form $m^{2k}T^D_{(2n,\pm 2n)}$ and $m^{2k}T^D_{(2,0)}$. 
In other words, the one-point functions $\vev{(\tfrac{\mu}{\t_2})^k \, T^D_{(2n,\pm2n)}}_{\tau,\mu;\a,\b}$ are modular forms of weight $(2n+k,k)$ and $(k,2n+k)$ respectively 
and $\vev{(\tfrac{\mu}{\t_2})^k \, T^D_{(2,0)}}_{\tau,\mu;\a,\b}$ are modular forms of weight $(k+1,k+1)$.

\subsection{Massless limit \label{sec:massless}}

In the massless limit the one point functions $\vev{T_{(2n,\pm 2n)}}_{\tau, 0; \a,\b}$ become holomorphic and anti-holomorphic functions of $\t$ respectively. They also correspond to the expectation value of the KdV charges in the $c=\tfrac{1}{2}$ minimal model. Since the other one point functions take the form $m^k\vev{T_{(2n,\pm 2n)}}_{\tau, m; \a,\b}$, for $k$ a positive even integer, they vanish in the massless limit so we only look at the edge cases.

We use the following limit for modified Bessel functions
\begin{align}
    \lim_{x\rightarrow0}x^{2n}K_{2n}(ax)=a^{-2n}2^{2n-1}(2n-1)!\;.
\end{align}
This gives us
\be
\begin{split}
    &\lim_{m\rightarrow 0}\vev{T_{(2n,2n)}}_{\tau,m;\a,\b}
    \=\frac{1}{2}(2n-1)!\sideset{}'\sum_{(r,\ell)\in\mathbb{Z}^2}\frac{\rme^{2\pi \ii(\a \ell+\b r)}}{(\t r-\ell)^{2n}}\;,\\
    &\lim_{m\rightarrow 0}\vev{T_{(2n,-2n)}}_{\tau,m;\a,\b}
    \= \frac{1}{2}(2n-1)!\sideset{}'\sum_{(r,\ell)\in\mathbb{Z}^2}\frac{\rme^{2\pi \ii(\a \ell+\b r)}}{(\bar\t r-\ell)^{2n}} \,,
\end{split}
\ee
which, by definition, are the holomorphic Kronecker-Eisenstein series. This is a known result for massless fermions. In order to relate these massless expressions to those in \cite{Downing:2021mfw} one can use the following definition for the Eisenstein series 
\begin{align}
    E_{2n}(\t) \=\frac{1}{2\zeta(2n)} \sideset{}'\sum_{(r,\ell)\in\mathbb{Z}^2}\frac{1}{(\t r + \ell)^{2n}}\;,
\end{align}
and the identities
\begin{align}
    &\sideset{}'\sum_{(r,\ell)\in\mathbb{Z}^2}\!\! \frac{(-1)^r}{(r\tau {+} \ell)^{2n}} {=} 2\zeta(2n)(2E_{2n}(2\t) {-} E_{2n}(\t))\,,\\
    &\sideset{}'\sum_{(r,\ell)\in\mathbb{Z}^2}\!\! \frac{(-1)^\ell}{(r\tau {+} \ell)^{2n}} {=} 2\zeta(2n)(2^{1-2n}E_{2n}(\tfrac{1}{2}\t) {-} E_{2n}(\t))\,,\\
    &\sideset{}'\sum_{(r,\ell)\in\mathbb{Z}^2}\!\! \frac{(-1)^{r+\ell}}{(r\tau {+} \ell)^{2n}} {=} 2\zeta(2n) \!(\!(1 {+} 2^{2-2n})E_{2n}(\t) {-} 2^{1-2n}E_{2n}(\tfrac{1}{2}\t) {-} 2E_{2n}(2\t)\!)\,.
\end{align}
We denote the KdV charges in the holomorphic sector by $I_{2n-1}$ and using the normalisation and notation in \cite{Downing:2021mfw} we have the relations\footnote{Note that~$\a=\b=0$ was not studied in~\cite{Downing:2021mfw} 
because the partition function vanishes for this spin sector in the massless limit.}
\begin{align}
    &\vev{T_{(2n,2n)}}_{\tau,0;0,\tfrac{1}{2}}=\frac{(-1)^{n+1}}{(2\pi)^{2n}}\vev{I_{2n-1}}^{\text{R},+}\;,\\
    &\vev{T_{(2n,2n)}}_{\tau,0;\tfrac{1}{2},0}=\frac{(-1)^{n+1}}{(2\pi)^{2n}}\vev{I_{2n-1}}^{\text{NS},-}\;,\\
    &\vev{T_{(2n,2n)}}_{\tau,0;\tfrac{1}{2},\tfrac{1}{2}}=\frac{(-1)^{n+1}}{(2\pi)^{2n}}\vev{I_{2n-1}}^{\text{NS},+}\;.
\end{align}
Analogous results hold for the anti-holomorphic sector.

\section{Differential equations obeyed by the one-point functions}
\label{sec:diffeq}
 
In this section we make comments on certain structural properties of the massive fermion one-point 
functions~$\vev{ T_{(2n,\pm 2n)}}_{\tau,\mu;\a,\b}$ with respect to the modular and Jacobi group. (Throughout we use the invariant mass $\mu$ introduced in \eqref{eq:invarmass}.)
In fact, we introduce a slightly more general family of functions in~\eqref{eq:Ftau} below which include these one-point functions
as well as the (logarithm of the) partition function~\eqref{defpartfnab} of the massive fermion theory. 
In the massless limit ($\mu \to 0$) these functions generalise the one-point functions and 
the logarithm of the partition function of massless free fermions, 
which are expressed in terms of classical Jacobi theta and Dedekind eta functions. The partition function~\eqref{defpartfnab} of the massive theory
has been discussed in the paper~\cite{Berg:2019jhh} where it is shown that 
it belongs to the framework of (Jacobi) Maass forms, and 
correspondingly obeys three differential equations. 
The first of these equations involves the hyperbolic Laplacian and the other two involve a Laplacian in the elliptic variable 
and the Casimir of the Jacobi group. 
We show below how the functions arising from the massive theory obey generalisations of these equations. 
Furthermore, we show that there are raising and lowering operators that move us through the members of the family. 

\medskip

Recall that the modular group acts on the variables~$(z,\t) \in \IC \times \IH$ as
\be \label{modtrans1}
(z,\t) \; \mapsto \; \Bigl( \, \frac{z}{c\t+d} \,  ,\, \frac{a\t+b}{c\t+d} \, \Bigr) \,, 
\qquad{\Bigl(\,{\small \begin{matrix}a&b\\c&d\end{matrix}}\,\Bigr)\in \slz } \,,
\ee
and the elliptic transformations of the Jacobi group acts as
\be \label{elltrans1}
z \; \mapsto \; z + \nu \t + \lambda \,, \qquad \nu, \lambda \in \IZ \,.
\ee
In order to discuss the differential equations, we introduce the following family of complex-valued 
functions on~$\IC \times \IH$, as sums over the lattice~$\Lambda_\t = \IZ \t + \IZ$,
\begin{align}\label{eq:Fztau}
    \CF^{n,m}_\mu(z,\tau) \=  \CF^{n,m}_\mu(h;z,\tau) \=
\sideset{}'\sum_{\omega \in \Lambda_\t} \frac{1}{\omega^n \, \overline{\omega}^m} \; 
    h \biggl(\mu\frac{\omega \overline{\omega}}{\tau_2} \biggr) \, \chi_{\omega}(z)\;,
\end{align}
where\footnote{More generally, $n,m$ can be half-integers or even real. 
In this case, one has to give a slightly more careful definition which takes into account 
potential branch cuts.} $n,m\in\mathbb{Z}$, 
and $h(x)$ is a real function which decays sufficiently fast at infinity so as to 
ensure convergence of the sum, and 
$\chi_{\omega}:z\mapsto \rme^{2\pi i\text{Im}(\omega \bar z)/\t_2}$
is a~$U(1)$ character on the torus. 
(The prime on the summation indicates that 
the origin of the lattice is removed from the sum.)
This series is of the Kronecker-Eisenstein type, and is not a holomorphic function of~$\t$ or of~$z$. 
The above expression as a lattice sum makes it manifest that~$\CF^{n,m}_\mu(z,\tau)$ is invariant under elliptic 
transformations of the form~\eqref{elltrans1} (i.e.~it has index~0) and transforms as a modular form of weight~$(n,m)$. 

Upon labelling the lattice points as~$\omega=r\t-\ell$, and writing~$z=\a\t+\b$ (so that~$\chi_{\omega}= \rme^{2\pi i(\alpha \ell+\beta r)}$), we obtain 
\be\label{eq:Ftau}
    \CF^{n,m}_\mu(\a\t+\b,\tau) \= 
\sideset{}'\sum_{(r,\ell) \in \Z^2} \frac{1}{(r\t-\ell)^n \, (r\bar\t-\ell)^m} \; 
    h \biggl(\mu\frac{|r\t-\ell|^2}{\tau_2} \biggr) \, e^{2\pi\\i(\a \ell+\b r)}\;.
\ee 
From the expression~\eqref{Tpmlatticesum} we see
that the one-point functions belong to this family (up to a prefactor). For~$\ell=1,2,\dots$, we have 
\be \label{TFrel}
\vev{ T_{(2\ell,\pm 2\ell)}} \= \Bigl(\frac{\mu}{4\t_2} \Bigr)^\ell 
\CF^{\pm \ell,\mp \ell}_\mu \bigl(h;z,\tau \bigr)  \quad \text{with~$h(x) = K_{2n}(\sqrt{x})$} \,.
\ee
As we see below, the functions~$\CF^{n,m}_\mu$ satisfy two
of the three differential equations referred to above 
(the hyperbolic Laplacian and the Casimir equation) for any~$h$,
while the third equation (the Laplacian in~$z$) is obeyed only for~$h$ being a $K$-Bessel function. 

Note that the functions~$ \CF^{n,m}_\mu$ are well-defined for~$n \in \IZ$ even though 
the operators~$T_{(2n,\pm 2n)}$ are only defined for~$n>0$.  

In the massless limit~$\mu \to 0$, the one-point functions for fixed values of~$\a$, $\b$ are holomorphic functions of~$\t$. When~$z=0$, these are the familiar holomorphic Eisenstein series, as discussed 
in Section~\ref{sec:massless}. 
The logarithm of the partition function of the massive free fermion~\eqref{defpartfnab} is also of this 
form with~$m=n=1/2$ and~$h$ being a~$K_1$-Bessel function as above~\cite{Berg:2019jhh}. 

In the following two subsections, we fix $\a$ and $\b$ and consider the functions \eqref{eq:Ftau} as a function of $\t$.

\subsection{Lowering and raising operators}
\label{ssec:lowerraise}

It is easy to check that the following differential operators 
\be \label{eq:raiselower}
    X_+ \= X_+^{(k)}\=(\t-\bar \t)\partial_\t + k\,, \qquad 
    X_- \= X_-^{(\bar k)}\=-(\t-\bar\t)\partial_{\bar\t} + \bar k\,,
\ee
map a modular form of weight $(k,\bar k)$ to one with weight $(k \pm 1,\bar k \mp 1)$. 
This leads us to label the operators $X_{\pm}$ as raising and lowering operators. In fact, these operators, together with 
the scaling operator
\be
    Z \= (k-\bar k)\,,
\ee
which again acts on modular forms of weight $(k,\bar k)$ as scalar multiplication, 
form an $\mathfrak{sl}_2(\R)$ algebra \cite{berndt2012elements},
\be\label{eq:sl2Rrep}
    [X_+,X_-] \= Z\,,\quad [Z,X_\pm] \= \pm 2X_\pm\,.
\ee
These operators factor through powers of $\t_2$ as follows,
\be
    X_\pm(\t_2^kf(\t,\bar\t)) \=\t_2^k X_\pm f(\t,\bar\t) \,, \qquad 
      Z(\t_2^kf(\t,\bar\t)) \= \t_2^k Z f(\t,\bar\t)\,.
\ee
The operators~$X_{\pm}$ relate the various one-point functions to each other. 
To see this, note that they act on the functions  $\CF^{n,m}_\mu(\a\t+\b,\t)$ as follows, 
\be \label{eq:raising}
\begin{split}
    &X_{+}\CF^{n,m}_\mu(\a\t+\b,\t)\=(-\mu\partial_\mu + n)\CF^{n+1,m-1}_{\mu}(\a\t+\b,\t)\,,\\
    &X_{-}\CF^{n,m}_\mu(\a\t+\b,\t)\=(-\mu\partial_\mu + m)\CF^{n-1,m+1}_{\mu}(\a\t+\b,\t)\,.
\end{split}
\ee
Using the relation~\eqref{TFrel}, we obtain, for~$n=0,1,2,\dots$, 
\be 
\begin{split}\label{eq:Traiselower}
&X_\pm\vev{ T_{(2n,\pm2n)}}_{\!\t,\mu;\a,\b} {=} -4\t_2\,\partial_\mu\vev{ T_{(2n+2,\pm(2n+2))}}_{\!\t,\mu;\a,\b}\;,\\
    &X_\mp\vev{ T_{(2n,\pm2n)}}_{\!\t,\mu;\a,\b} {=} \frac{1}{4\t_2}(-\mu^2\partial_\mu {+} 2\mu(n-1))\vev{ T_{(2n-2,\pm(2n-2))}}_{\!\t,\mu;\a,\b}\;.
\end{split}
\ee
We can therefore move along the two edges of the cone 
described in  Figure~\ref{fig:tensor pyramid} using the $X_\pm$ operators, thus relating all the one-point functions 
with purely holomorphic weight to each other. (Similarly, the one-point functions with purely 
anti-holomorphic weight are all related to each other). 
Notice that the first and second lines of \eqref{eq:Traiselower} also hold
for~$n=0$ and~$n=1$, respectively, and this 
allows us to go around the apex of the cone in Figure~\ref{fig:tensor pyramid}, thus relating 
all one-point functions with holomorphic or anti-holomorphic weight. 

Furthermore, we can relate all these one point functions to the partition function. In \cite{Berg:2019jhh} the following expression for the partition function is derived,
\be
    \log Z_{\a,\b} (\t,\taubar) = -\frac{\sqrt{\mu \t_2}}{\pi} \sideset{}'\sum_{(\ell,r)\in\Z^2} \frac{1}{|\t r - \ell|}K_1 \!\left(\! \sqrt{\frac{\mu}{\t_2}} |\t r - \ell| \!\right)\! e^{2\pi \ii (\a \ell + \b r)}\;,
\ee
where the prime indicates as before that we exclude the origin from the sum. One can explicitly see the action of the differential operators:
\be\label{eq:XlogZ}
    X_\pm \log Z_{\a,\b} \= -\frac{2\t_2}{\pi} \vev{T_{(2,\pm2)}}_{\t,\mu;\a,\b}\;,
\ee

\be\label{eq:dmlogZ}
    \partial_\mu \log Z_{\a,\b} \= \frac{1}{2\pi}\vev{T_{(0,0)}}_{\t,\mu;\a,\b} \,,
\ee
as mentioned at the beginning of Section~\ref{sec:massiveferm}.

In the massless limit, the one-point functions are purely holomorphic (anti-holomorphic) Kronecker-Eisenstein series, and therefore are annihilated by~$X_-$ ($X_+$). 

\subsection{Massive generalisation of (harmonic) Maass forms}
\label{ssec:MF}

The paper~\cite{Berg:2019jhh} introduces a notion of \emph{massive Maass forms} as follows. 
Introduce the hyperbolic Laplacian of weight $(k,\bar k)$ which acts on a modular form of weight $(k,\bar k)$
\be \label{DelXXrel}
\begin{split}
    \Delta_{\tau,k,\bar k}&\=-\frac{1}{2}(X_+X_-+X_-X_+)\\
    &\=(\t-\bar\t)^2\partial_\t\partial_{\bar\t}+(\t-\bar\t)(k\partial_{\bar\t}
    -\bar k\partial_{\t})+\tfrac{1}{2}(k+\bar k-2k\bar k) \,.
\end{split}
\ee
This Laplacian factors through powers of~$\t_2$ as follows,
\be
    \Delta_{\tau,k-n,\bar k-n}(\t_2^nf(\t,\bar\t))=\t_2^n \Delta_{\tau,k,\bar k}f(\t,\bar\t)\;.
\ee
According to Definition 1 in \cite{Berg:2019jhh}, a function $f(\mu,\t,\bar\t)$ is 
called a massive Maass form of weight $(k,\bar k)$ if for fixed $\mu$ it transforms as a modular 
form of weight $(k,\bar k)$ and satisfies the partial differential equation
\begin{align} \label{MMaassDE}
    \Delta_{\t,k,\bar k}f=(g_2(\mu)\partial_\mu^2+g_1(\mu)\partial_\mu+g_0(\mu))f\;,
\end{align}
for some smooth functions $g_i(\mu)$.

\medskip

Using the expression~\eqref{DelXXrel} for the Laplacian in terms of the raising and lowering operators, 
and their action~\eqref{eq:raising}, 
it is easy to check that the functions $\CF^{n,m}_\mu(\a\t+\b,\t)$ satisfy~\eqref{MMaassDE}
with
\be
g_2(\mu) \= - \mu^2 \,, \quad g_1(\mu) \=  \mu(n+m-2)  \quad g_0(\mu) \= \tfrac{1}{2}(n+m-2nm) \,,
\ee

i.e., they are weight $(n,m)$ massive Maass forms. Note that for the special cases $m=-n$, which are related to the one point functions the equation simplifies to
\begin{align}
\Delta_{\t,n,-n} \, \CF^{n,-n}_\mu(\a\t+\b,\t)\=
(-\mu^2\partial_\mu^2 - 2\mu\partial_\mu+n^2)\CF^{n,-n}_\mu(\a\t+\b,\t) \,,
\end{align}
By Lemma 3.1 in \cite{Berg:2019jhh} the function~$\mu^k \CF^{n,m}_\mu(\a\t+\b,\t)$
is also a massive Maass form of weight $(n,m)$ for any $k$, hence the one point 
functions $\vev{ T_{(2n,\pm 2n)}}_{\tau,\mu;\a,\b}$ are all massive Maass forms.

In the massless limit the one-point functions become either holomorphic or anti-holomorphic in $\t$. 
In both these cases, the second line of~\eqref{DelXXrel} makes it clear that the hyperbolic Laplacian acts 
trivially on the one point functions as multiplication by half the modular weight. 

\subsection{Differential equations arising from the Jacobi group}

\label{ssec:Jg}

As was shown in~\cite{Berg:2019jhh}, the partition function of the massive fermion theory
fits into the structure of Jacobi Maass forms as presented in \cite{Pitale2009}. Here we show that the 
one-point functions also fit into the same structure. 

In this subsection we return to using~$z$ as a variable. 
The operators $X_\pm$ defined in \eqref{eq:raiselower} are written as 
\be\label{eq:Xdiffop}
    X_+ = (\t-\bar\t)\partial_\t + (z-\bar z) \partial_z + k\,, \quad 
    X_- = -(\t-\bar\t)\partial_{\bar\t} - (z-\bar z) \partial_{\bar z} + \bar k\,.
\ee
We also introduce the derivatives 
\be\label{eq:Ydiffop}
    Y_{+} \= i\sqrt{\frac{\t-\bar\t}{2\ii}}\, \partial_z\;,\quad Y_- \= -i\sqrt{\frac{\t-\bar\t}{2\ii}}\, \partial_{\bar z} \,,
\ee
which map a modular form of weight $(k,\bar k)$ to one of weight $(k \pm \tfrac{1}{2}, \bar k \mp \tfrac{1}{2})$. 
Acting on $\CF^{n,m}_\mu(z,\t)$ with the $Y_\pm$ operators gives
\be
\begin{split}
    &Y_+\CF^{n,m}_\mu(z,\t)\=-\frac{i\pi}{\sqrt{\t_2}}\CF^{n,m-1}(z,\t)\;,\\
    &Y_-\CF^{n,m}_\mu(z,\t)\=-\frac{i\pi}{\sqrt{\t_2}}\CF^{n-1,m}(z,\t)\;.
\end{split}
\ee
We now define the Laplacian in the~$z$ space
\begin{align}
    \Delta_z =-i(\t-\bar\t)\partial_z\partial_{\bar z} \,,
\end{align}
and the Casimir operator
\begin{align}
    C_{\t,z,k,\bar k} = \frac{\t{-}\bar\t}{2i} \!\left(\! (\t{-}\bar\t)(\partial_{\bar\t}\partial_z^2
     {+} \partial_\t\partial_{\bar z}^2) {+} (z{-}\bar z)(\partial_{\bar z}\partial_z^2 {+} \partial_z\partial_{\bar z}^2)
    {+} (\partial_z {+} \partial_{\bar z})(k\partial_{\bar z} {-} \bar k\partial_z)\!\right)\! \,.
\end{align}
As in \cite{berndt2012elements} these operators can be written in terms of the $X_\pm$ and $Y_\pm$ above as
\begin{align}
    &\Delta_z = Y_+Y_- + Y_-Y_+\,,\label{eq:z Laplace}\\
    &C_{\t,z,k,\bar k} =\! \frac{1}{2}(Y_+^2X_- +\! X_-Y_+^2 -\! Y_-^2X_+ -\! X_+Y_-^2 + \!(k {-} \bar k)\!(Y_+Y_- +\! Y_-Y_+ \!)\!). \label{DelCdef}
\end{align} 
These operators factor through powers of~$\t_2$ as
\begin{align}
    &\Delta_z(\t_2^kf(z,\t)) \=\t_2^k\Delta_zf(z,\t)\;, \\
    &C_{\t,z,k-n,\bar k-n}(\t_2^n f(z,\t)) \=\t_2^nC_{\t,z,k,\bar k}f(z,\t)\;.
\end{align}
We now consider the definition of a \emph{massive Maass-Jacobi form} as given in \cite{Berg:2019jhh}
and apply it to our case with index~0. 
We say that a function $f(\mu,\t,\bar\t,z,\bar z)$ is a \emph{massive 
Maass-Jacobi form} of index~$0$ and weight $(k,\bar k)$ if for fixed $\mu$ it transforms as a Jacobi form 
of index~0 and weight $(k,\bar k)$ and satisfies the partial differential equations
\begin{align} 
   & C_{\t,z,k,\bar k}f \=0\;, \label{MMaasJac1}\\
    &\Delta_z f \= (G_2(\mu)\partial_\mu^2 + G_1(\mu)\partial_\mu + G_0(\mu))f\;, \label{MMaasJac2}
\end{align}
for some smooth 
functions~$G_n(\mu)$.\footnote{The paper~\cite{Berg:2019jhh} also make a remark about higher order 
differential operators on the right-hand side of~\eqref{MMaasJac2}, but we do not discuss this here.}
This definition is given in~\cite{Berg:2019jhh} for massive Jacobi-Maass 
forms of general index $(m,\overline{m})$ using a suitable modification of the differential 
operators $C_{\t,z,k,\bar k}$ and $\Delta_z$ above, but we will not need this in our discussion. 

\medskip

Now we discuss the family~$\CF^{n,m}_\mu(z,\t)$ in the above context. 
Using the expression~\eqref{DelCdef}, it is easy to check that the 
functions $\CF^{n,m}_\mu(z,\t)$ are annihilated by the Casimir operator
\begin{align}
    C_{\t,z,n,m}\CF^{n,m}_\mu(z,\t)\=0 \,.
    \label{eq:Casimir}
\end{align}
The condition that~$\CF^{n,m}_\mu(z,\t)$ satisfies~\eqref{MMaasJac2} implies that 
there are functions $G_n(\mu)$ with 
\begin{align}
    \Delta_{z}\CF^{n,m}_\mu(z,\t) \= (G_2(\mu)\partial_\mu^2 + G_1(\mu)\partial_\mu + G_0(\mu))\CF^{n,m}_\mu(z,\t)\;.
\end{align}
This is equivalent to the function $h(x)$ satisfying the following differential equation
\be
    2\pi^2 x h(x) + \mu G_0(\mu) h(x) + x G_1(\mu)h^{\prime}(x) + \frac{x^2}{\mu}G_2(\mu)h^{\prime\prime}(x) \=0\;.
\ee
Since the first term is independent of~$\mu$ for every~$x$, the same should be true for all the terms which forces the $G_n$ to all take the form
\be
    G_n(\mu)\=(2\pi)^2\lambda_n \, \mu^{n-1}\;,
\ee
where $\lambda_n$ are constants. This gives us the differential equation
\be\label{eq:hdiffeq}
    (x + \lambda_0) h(x) + \lambda_1 x h^{\prime}(x) + \lambda_2 x^2 h^{\prime\prime}(x)\=0\;.
\ee
If the function $h(x)$ satisfies the above differential equation, the function $\CF^{n,m}_\mu(h;z,\t)$ satisfies
\be
    \Delta_{z}\CF^{n,m}_\mu(z,\t)\=
    (2\pi)^2\left(\lambda_2\, \mu\, \partial_\mu^2 + \lambda_1\, \partial_\mu + \frac{\lambda_0}{\mu}\right)\CF^{n,m}_\mu(z,\t)\;.\label{eq:z laplacian}
\ee
The above differential equation can be rewritten, in the variable $y=2\ii\sqrt{{x}/{\lambda_2}}$ and upon setting~$f(y)=\left(\tfrac{i}{2}\sqrt{\lambda_2}\,y \right)^{\frac{\lambda_1-\lambda_2}{\lambda_2}}h\left(-\tfrac14 \lambda_2y^2 \right)$, as
\begin{align}
    y^2 f^{\prime\prime}(y) + y f^\prime(y) - (y^2 + \nu^2)f(y) \=0\;,\quad \nu^2 \= \left(\frac{\lambda_1}{\lambda_2}-1\right)^2-\frac{4\lambda_0}{\lambda_2} \,.
\end{align}
The above equation is the modified Bessel equation whose two linearly independent solutions 
are~$I_\nu(y)$ and $K_\nu(y)$, the modified Bessel functions of the first and second kind respectively. 
Since the function $I_{\nu}(y)$ grows exponentially and $K_\nu(y)$ decays exponentially as $y \to \infty$ (for $y \in \R$), the solution to \eqref{eq:hdiffeq} 
can only contain $K_\nu$ for convergence of the lattice sum~\eqref{eq:Fztau}.
Note that for~$\lambda_2=\lambda_1=-4$ and $\lambda_0=(2n)^2$, 
the exponentially decaying solution to~\eqref{eq:hdiffeq} is~$h(x)=K_{2n}(\sqrt{x})$,
which is precisely the value (see~\eqref{TFrel}) for which $\CF^{n,m}_\mu(z,\t)$ is related to the one point functions of the massive fermions. 

\medskip

It is also interesting to note the relation of the mass term~$m^2\t_2=\mu$ to this Casimir operator. 
If we replace the parameter $\mu$ with $\mu f(\tau,\bar\tau)$ where the function $f(\t,\bar\t)$ 
is a weight $(0,0)$ modular form then the function $\CF^{n,m}_{\mu f}(z,\t)$ is again a 
Jacobi form of index~0 and weight $(n,m)$. 
Now, applying the Casimir to the series definition of~$\CF^{n,m}_{\mu f}$ shows that it is only annihilated 
if $f(\t,\bar\t)$ satisfies the partial differential equation
\begin{align}
    (\bar\t r-\ell)^2\partial_{\bar\t}f+(\t r-\ell)^2\partial_{\t}f\=0 \,,
\end{align}
for every~$r,\ell \in \IZ$. 
Clearly, the only solution is $f=$ constant. 
This means that if we want the one point function to be annihilated by 
the Casimir the only option for the mass term is~$m^2\t_2=\mu$.

\section{Conclusions}
\label{sec:conc}

On physical grounds, the torus one-point expectation values of fields in massive quantum field theories should have well-defined modular properties. We have calculated these for a range of tensor fields in the massive free fermion model and shown that, upon a rescaling of the mass, these are indeed modular forms. In addition, these one point functions obey three differential equations that arise from the representation theory of the Jacobi group. 

In the massless limit,
the fields $T_{(2n,\pm 2n)}$ we have considered become chiral fields (for $n>0$), 
but it is well known that one point functions of non-chiral fields also have well-defined modular properties
\cite{Moore-Seiberg,Gaberdiel_2009}, as do multi-point functions of chiral fields \cite{Zhu1996}.
Obvious further ideas are to try to extend our results to similar cases in the massive free fermion/Ising model.

All these ideas could be extended, in principle, 
to other massive field theories. However, since there are few examples where the partition functions 
and correlation functions are known explicitly, this may prove challenging.
More generally one would like to obtain a mathematical proof that such modularity properties must exist in massive theories (in the way that \cite{Zhu1996} gives a general proof for rational conformal field theories). \cite{Zhu1996} uses the underlying algebraic structure when proving modular invariance and hence if one wished to proceed in a similar fashion then the underlying algebra of the massive theories would have to be well understood.

Most immediately, this project was, in part, motivated by the desire to analyse generalised Gibbs ensembles in the massive free fermion model, extending the work in \cite{Downing:2021mfw}. Since the one-point functions do indeed have the same modular properties as those in the massless CFT it is very likely that the GGE in the massive case will have similar modular properties to those discussed in \cite{Downing:2021mfw}. We hope to return to this in the future.

\subsection*{Acknowledgments}

We would like to thank B.~Doyon, J.~Harvey, and I.~Runkel for comments and discussions.
We would also like to thank M.~Berg for discussions, comments on the manuscript and sharing a draft version of \cite{Berg}.
This work was supported by the EPSRC grant EP/V520019/1, the  ERC Consolidator Grant N. 681908, “Quantum black holes: A microscopic window into
the microstructure of gravity”, and the STFC grant ST/P000258/1.

\appendix

\section{One-point functions and modular forms \label{sec:PImodform}}

In this appendix we explain the functional integral argument for why the one-point function of any tensor field in a 
translation-invariant theory on the torus is a modular form.

By a torus of modular parameter~$\t \in \IH$, we mean the elliptic 
curve~$E_\t=\IC/(\IZ\t+\IZ)$. We use the complex coordinates~$x=(x^1,x^2) = (z,\zbar)$ on the torus. 
The modular group has an action on the modular parameter as 
\be \label{modtrans}
\gamma: \t \; \mapsto \; \frac{a\tau+b}{c\tau+d} \; \equiv \; \wh\t \,, 
\qquad \gamma \= \Big(\,{\small \begin{matrix}a&b\\c&d\end{matrix}}\,\Big)\in \slz \,.
\ee
The action \eqref{modtrans} relates equivalent tori. This equivalence is implemented by the map 
\be \label{coordtrans}
(z,\zbar) \; \to \; \bigl(\,\wh z \,,\, \wh \zbar \, \bigr) \= 
 \Bigl(\, \frac{z}{c\tau+d} \,,  \frac{\zbar}{c\taubar+d} \, \Bigr) \,,
\ee
which maps the torus $E_\tau$ to $E_{\hat\tau}$. If 
$f_{\hat \tau}$
is a well-defined function on $E_{\hat \tau}$, 
then $f_\tau(z,\bar z)=f_{\hat \tau}(\hat z,\hat{\bar z})$ is well-defined on $E_\tau$. 
We can say that $f_\tau(z,\bar z)$ is invariant under the simultaneous transformations~\eqref{modtrans},~\eqref{coordtrans}.

The measure
\be \label{dmumeasure}
d\mu \= \frac{\dd z \,\dd\zbar}{\Im(\t)}\;,
\ee
is also invariant under such a simultaneous transformation.

\bigskip

We want to study the properties of the function $\tau \mapsto \vev{T}_\tau$ where $T$ is a tensor field. 
A tensor field~$T_{\mu_1\dots\mu_n}(x)$ transforms, under a transformation~$x \mapsto \wh x$  as
\be\label{tensortrans}
T_{\mu_1\dots\mu_n} (x) \; \to \; 
\frac{\partial x^{\nu_1}}{\partial \wh x^{\mu_1}}\dots\frac{\partial x^{\nu_n}}{\partial \wh x^{\mu_n}} \,
T_{\nu_1\dots\nu_n} (x) \,.
\ee
Note that the coordinate transformation~\eqref{coordtrans} acts on~$z,\zbar$ separately, so that the
components~$T_{\mu_1\dots\mu_n}$ of the tensor diagonalise the action of this transformation on the tensor field.
(In other words, under the transformation~\eqref{coordtrans}, the tensor components on the two sides 
of~\eqref{tensortrans} have the same indices.) 
In what follows, we focus on transformations that consist of a combination of rotations and scaling transformations, i.e.,
\be
(z,\zbar) \; \to \; \bigl(z\, \rme^{\ii \theta}, \zbar \, \rme^{-\ii \theta}\bigr) \,, \qquad 
(z,\zbar) \; \to \;  (r z, r \zbar)\,, \qquad r, \theta \in \IR \,,
\ee
or, more compactly,
\be \label{zscaling}
(z,\zbar) \; \to \; \bigl(\lambda z, \overline{\lambda} \zbar \bigr) \,, \qquad \lambda \in \IC \,.
\ee
A component of the tensor field $T_{\mu_1\dots\mu_n}$ with~$m$ of its legs~$\mu_i=z$ 
and~$n-m$ of them with $\mu_i=\zbar$ 
transforms under~\eqref{zscaling} as 
\be \label{tenstransscaling}
T_{\mu_1\dots\mu_n} \bigl(z, \zbar \bigr) \; \to \; \lambda^{-m} \, {\overline{\lambda}}\vphantom{\lambda}^{m-n} \,
T_{\mu_1\dots\mu_n} \bigl(z, \zbar \bigr) \,.
\ee
Now we consider a quantum theory of fields (generically denoted by~$\phi$ below) 
defined on the torus~$E(\t)$ with an action functional 
\be\label{Lagrangian}
S_\tau[\phi] \= \int_{E_\t} \, d\mu \, \Im(\t) \, \CL \bigl(\phi (z,\zbar) \bigr)\,,
\ee
where~$\CL$ is the Lagrangian density. 
The quantum expectation value of any operator $\mathcal{O}(\phi)$ in the QFT is defined by the path integral
\begin{align} \label{Oonept}
\vev{\mathcal{O}(\phi)}_\tau \= \frac{1}{Z_\tau} \int \, [D\phi] \, \mathcal{O}(\phi) \, \rme^{-S_\tau[\phi]} \,,
\end{align}
where 
\be \label{defZPI}
Z_\tau \= \int \, [D\phi] \, \rme^{-S_\tau[\phi]} \,,
\ee
is the partition function of the theory.

A theory which is invariant under the symmetry transformation discussed above obeys
\be
S_\tau[\phi] \= S_{\wh \tau}[\wh \phi] \,.
\ee
The meaning of the right-hand side of the above equation is that we transform the field 
$\phi(z,\zbar) \to \wh \phi (\wh z, \wh \zbar)$ according to the tensorial properties of the field~$\phi$ 
under~\eqref{modtrans},~\eqref{coordtrans}, 
and integrate the resulting Lagrangian over~$(\wh z, \wh \zbar)$ coordinates.
When this symmetry is not anomalous in the quantum theory, the  measure $[D\phi]$ of the 
path integral~\eqref{defZPI} is also invariant under the transformation. As a consequence, the partition function 
is invariant under the symmetry transformation, i.e.,
\be
Z_\tau \= Z_{\wh \tau} \,.
\ee
In such a theory, consider a local tensor operator~$T_{\nu_1\dots\nu_n}(x)$ transforming as~\eqref{tensortrans}.
Its quantum expectation value~$\vev{T_{\nu_1\dots\nu_n} (x)}_\tau$, as defined by~\eqref{Oonept},
transforms under the symmetry transformation~\eqref{modtrans},~\eqref{coordtrans} as
\be \label{Ttrans}
\begin{split}
\vev{T_{\mu_1\dots\mu_n} (x)}_{\tau}
& \; \to \; 
\frac{1}{Z_\t}\int \, [D\phi] \, \frac{\partial x^{\nu_1}}{\partial \wh x^{\mu_1}}\dots\frac{\partial x^{\nu_n}}{\partial \wh x^{\mu_n}} \,
T_{\nu_1\dots\nu_n} (x)\, \rme^{- S_{\tau}[\phi]} \\
& \=  \frac{\partial x^{\nu_1}}{\partial \wh x^{\mu_1}}\dots\frac{\partial x^{\nu_n}}{\partial \wh x^{\mu_n}} \,
\vev{T_{\nu_1\dots\nu_n} (x)}_\tau \,.
\end{split}
\ee
Now we consider a theory of the above sort that is also translation-invariant.
When the theory is translation-invariant, the one-point function~$\vev{T_{\mu_1\dots\mu_n}}_\t$ is independent of~$x$. 
In such a theory, consider a component of the tensor field $T_{\mu_1\dots\mu_n}$ with~$m$ of its legs~$\mu_i$ equal to $z$ 
and~$n-m$ of them equal to~$\zbar$.
The one-point function of this tensor field~$\vev{T_{\mu_1\dots\mu_n}}_\tau$ is a function only of~$\t,\taubar$.
Since the coordinate transformation~\eqref{coordtrans} 
is a combination of rotation and scaling with~$\lambda = (c\t+d)^{-1}$,
the transformation~\eqref{tenstransscaling} implies that under~$\t \to \wh \t$, 
\be \label{Ttransmodular}
\vev{T_{\mu_1\dots\mu_n}}_{\tau} \; \to \;
  (c\tau+d)^m \, (c\bar\tau+d)^{n-m} \,
\vev{T_{\mu_1\dots\mu_n}}_\tau \,,
\ee
i.e., the function~$\t \mapsto \vev{T_{\mu_1\dots\mu_n}}_\tau$ 
is a non-holomorphic modular form of weights $(m,n-m)$. 
Any CFT automatically satisfies the symmetry requirements that we assumed above, and therefore gives rise to modular forms.

\medskip

Now we apply this general discussion to the theory of massive fermions. 
Consider the Lagrangian density for Euclidean Majorana free fermions
\be  \label{Lagpsipsibar}
 \CL \=   \CL_0 +  m\Psi\bar\Psi \,, \qquad 
  \CL_0\= \Psi\bar\partial\Psi - \bar\Psi \,\partial\bar\Psi\;.
\ee 
Under the symmetry transformation \eqref{zscaling} the fermion fields $\Psi$ and $\bar\Psi$ 
have the transformation property
\be
    \Psi(z,\bar z) \; \to \; \lambda^{-\frac{1}{2}}\Psi(z, \bar z) \;,\qquad 
    \bar\Psi(z,\bar z) \; \to \; \bar\lambda^{-\frac{1}{2}}\bar\Psi(z, \bar z)\;,
\ee
and the derivatives transform as
\be
    \partial \; \to \; \lambda^{-1}\partial \;,\qquad \bar\partial \; \to \; \bar\lambda^{-1}\bar\partial\;.
\ee
This implies that the kinetic terms scale with weight~2 under the transformation \eqref{zscaling}, i.e.,
\be
\CL_0\bigl(\phi(z, \zbar)\bigr) \= |\lambda|^{-2} \, \CL_0\bigl(\phi(\lambda z,\bar\lambda \zbar)\bigr) \,.
\ee
Indeed, the above scaling with a weight of~2 is obeyed by any two-dimensional conformal field theory with a Lagrangian 
description.

Now, the mass term in the Lagrangian~\eqref{Lagpsipsibar} for fixed mass scales as~$1/|\lambda|$, but 
it scales as~$1/|\lambda|^2$ if the transformation on the fields is accompanied by the following transformation 
of the mass parameter  
\be\label{eq:mtransform}
    m \; \to \;  m/|\lambda| \;.
\ee
It is therefore natural to define~$\mu = m^2 \Im(\tau)$, which is kept fixed under the modular transformation. 
Under this condition, the one-point functions in the massive fermion theory transforms exactly as~\eqref{Ttransmodular}.

\section{Quantising the free fermion}\label{Quantising fermion}
In this appendix we outline the conventions used for the fermion fields and define what we mean by an Euclidean Majorana fermion.

\subsection{Dirac fermions}
We start with a two component Lorentzian, Dirac fermion, $(\Psi,\bar\Psi)$, satisfying the equations of motion
\begin{align}\label{eq:fermioneom}
    &(\partial_x-\partial_t)\Psi=-m\bar\Psi\;,\\
    &(\partial_x+\partial_t)\bar\Psi=-m\Psi\;,
\end{align}
and impose the periodicity conditions
\begin{align}
    \Psi(x+R,t)=e^{2\pi i \a}\Psi(x,t)\;,\;\bar\Psi(x+R,t)=e^{2\pi i \a}\bar\Psi(x,t)\;,
\end{align}
for any $\a$, on the fields. The mode expansion for the fields is
\begin{align}
&\bar\Psi(x,t)=\sqrt{\frac{2\pi}{R}}\sum_{k\in\mathbb{Z}+\alpha} \!\left(\! \sqrt{\tfrac{\omega_k^-}{2\omega_k}}\Psi_ke^{\frac{i\pi}{4}}e^{i(\omega_k t+\frac{2\pi k}{R}x)}+\sqrt{\tfrac{\omega_k^+}{2\omega_k}}\chi_k^\dagger e^{-\frac{i\pi}{4}}e^{-i(\omega_k t-\frac{2\pi k}{R}x)} \!\right),\\
&\Psi(x,t)=\sqrt{\frac{2\pi}{R}}\sum_{k\in\mathbb{Z}+\alpha} \!\left(\! \sqrt{\tfrac{\omega_k^+}{2\omega_k}}\Psi_k e^{-\frac{i\pi}{4}}e^{i(\omega_k t+\frac{2\pi k}{R}x)}+\sqrt{\tfrac{\omega_k^-}{2\omega_k}}\chi_k^\dagger e^{\frac{i\pi}{4}}e^{-i(\omega_k t-\frac{2\pi k}{R}x)} \!\right),
\end{align}
where
\begin{align}\label{eq:omega}
    \omega_k=\sqrt{m^2+\tfrac{4\pi^2 k^2}{R^2}}\;,\;\omega_k^{\pm}=\omega_k\pm\tfrac{2\pi k}{R}\;.
\end{align}
These fields are not hermitian and hence in addition we need the equations of motion for $(\Psi^\dagger,\bar\Psi^\dagger)$, give by
\begin{align}
    &(\partial_x-\partial_t)\Psi^\dagger=-m\bar\Psi^\dagger\;,\\
    &(\partial_x+\partial_t)\bar\Psi^\dagger=-m\Psi^\dagger\;.
\end{align}
Canonical quantisation gives the mode anti commutators
\begin{align}
    \{\Psi_n^\dagger,\Psi_m\}=\delta_{n,m}\;,\quad\{\chi_n^\dagger,\chi_m\}=\delta_{n,m}\;.
\end{align}
If we Wick rotate the $t$ coordinate to be $t=iy$ and define the complex coordinates $z=x+iy$ and $\bar z=x-iy$ the mode expansion of the fields is
\begin{align}
&\bar\Psi(z,\bar z)=\sqrt{\frac{2\pi}{R}}\sum_{k\in\mathbb{Z}+\alpha} \!\left(\! \sqrt{\tfrac{\omega_k^-}{2\omega_k}}\Psi_k e^{\frac{i\pi}{4}}e^{\tfrac{i}{2}(\omega_k^+z-\omega_k^-\bar z)}+\sqrt{\tfrac{\omega_k^+}{2\omega_k}}\chi_k^\dagger e^{-\frac{i\pi}{4}}e^{-\tfrac{i}{2}(\omega_k^-z-\omega_k^+\bar z)}\!\right),\\
&\Psi(z,\bar z)=\sqrt{\frac{2\pi}{R}}\sum_{k\in\mathbb{Z}+\alpha} \!\left(\! \sqrt{\tfrac{\omega_k^+}{2\omega_k}}\Psi_k e^{-\frac{i\pi}{4}}e^{\tfrac{i}{2}(\omega_k^+z-\omega_k^-\bar z)}+\sqrt{\tfrac{\omega_k^-}{2\omega_k}}\chi_k^\dagger e^{\frac{i\pi}{4}}e^{-\tfrac{i}{2}(\omega_k^-z-\omega_k^+\bar z)}\!\right),
\end{align}
The periodicity condition is now expressed as
\begin{align}
    \Psi(z+R,\bar z+R)=e^{2\pi i\a}\Psi(z,\bar z)\;,\;\bar\Psi(z+R,\bar z+R)=e^{2\pi i\a}\bar\Psi(z,\bar z)\;.
\end{align}
The equations of motion are
\begin{align}
    &\partial\bar\Psi=-\tfrac{1}{2}m\Psi\;,\quad \partial\bar\Psi^\dagger=-\tfrac{1}{2}m\Psi^\dagger\;,\\
    &\bar\partial\Psi=-\tfrac{1}{2}m\bar\Psi\;,\quad \bar\partial\Psi^\dagger=-\tfrac{1}{2}m\bar\Psi^\dagger\;.\\
\end{align}
These formally come from the Lagrangian density
\begin{align}\label{eq:DiracLag}
    \mathcal{L}=\Psi^\dagger\bar\partial\Psi-\bar\Psi^\dagger\partial\bar\Psi+\frac{1}{2}m(\Psi^\dagger\bar\Psi-\bar\Psi^\dagger\Psi).
\end{align}
The Lagrangian density is invariant under a coordinate rotation given by
\be
    (z,\bar z)\rightarrow(e^{i\theta}z,e^{-i\theta}\bar z)\;,
\ee
if the fields transform as
\begin{align}
    &\Psi(z,\bar z)=e^{\frac{i\theta}{2}}\Psi(e^{i\theta}z,e^{-i\theta}\bar z)\;,\;\Psi(z,\bar z)^\dagger=e^{\frac{i\theta}{2}}\Psi(e^{i\theta}z,e^{-i\theta}\bar z)^\dagger\;,\\
    &\bar\Psi(z,\bar z)=e^{-\frac{i\theta}{2}}\bar\Psi(e^{i\theta}z,e^{-i\theta}\bar z)\;,\;\bar\Psi(z,\bar z)^\dagger=e^{-\frac{i\theta}{2}}\bar\Psi(e^{i\theta}z,e^{-i\theta}\bar z)^\dagger\;.
\end{align}
Hence bilinear fields of the form $\Psi^\dagger\partial^{2n-1}\Psi$ and $\bar\Psi^\dagger\bar\partial^{2n-1}\bar\Psi$ transform under coordinate rotations as
\begin{align}
    &\Psi^\dagger\partial^{2n-1}\Psi\rightarrow e^{-2n\ii\theta}\Psi^\dagger\partial^{2n-1}\Psi\;,\\
    &\bar\Psi^\dagger\bar\partial^{2n-1}\bar\Psi\rightarrow e^{2n\ii\theta}\bar\Psi^\dagger\bar\partial^{2n-1}\bar\Psi\;.
\end{align}

\subsection{Majorana fermions}
\label{ssec:Mf}

We again solve the equations of motion \eqref{eq:fermioneom} but now impose a Majorana condition. The mode expansion is
\begin{align}
&\bar\Psi(x,t) {=} \sqrt{\frac{2\pi}{R}} \!\sum_{n\in\mathbb{Z}+\alpha} \!\!\sqrt{\tfrac{\omega_n^-}{2\omega_n}} \!\left(\! \Psi_ne^{\frac{i\pi}{4}}e^{i(\omega_n t+\frac{2\pi n}{R}x)} {+} \Psi_n^\dagger e^{-\frac{i\pi}{4}}e^{-i(\omega_n t+\frac{2\pi n}{R}x)} \!\right),\\
&\Psi(x,t) {=} \sqrt{\frac{2\pi}{R}} \!\sum_{n\in\mathbb{Z}+\alpha} \!\!\sqrt{\tfrac{\omega_n^+}{2\omega_n}} \!\left(\! \Psi_ne^{-\frac{i\pi}{4}}e^{i(\omega_n t+\frac{2\pi n}{R}x)} {+} \Psi_n^\dagger e^{\frac{i\pi}{4}}e^{-i(\omega_n t+\frac{2\pi n}{R}x)} \!\right),
\end{align}
where now $\alpha\in\{0,\tfrac{1}{2}\}$ and we define $\omega_n$ and $\omega_n^\pm$ as in \eqref{eq:omega}. These fields again have the periodicity property
\begin{align}
    \bar\Psi(x+R,t)=e^{2\pi i\a}\bar\Psi(x,t)\;,\;\Psi(x+R,t)=e^{2\pi i\a}\Psi(x,t)\;,
\end{align}
and we must impose $\alpha\in\{0,\tfrac{1}{2}\}$ in order to respect the reality condition on the fermions. After canonical quantisation the fermion modes have the anti commutation relations
\begin{align}
    \{\Psi_k,\Psi_l^\dagger\}=\delta_{k,l}\;.
\end{align}
If we Wick rotate the $t$ coordinate to be $t=iy$ and define the complex coordinates $z=x+iy$ and $\bar z=x-iy$ we get the fields
\begin{align}
&\bar\Psi(z,\bar z) {=} \sqrt{\frac{2\pi}{R}} \!\!\sum_{n\in\mathbb{Z}+\alpha}\!\!\! \sqrt{\tfrac{\omega_n^-}{2\omega_n}} \!\!\left(\!\! \Psi_ne^{\frac{i\pi}{4}}e^{\tfrac{i}{2}(\omega_n^+z-\omega_n^-\bar z)} {+} \Psi_n^\dagger e^{-\frac{i\pi}{4}}e^{-\tfrac{i}{2}(\omega_n^+z-\omega_n^-\bar z)} \!\!\right),\\
&\Psi(z,\bar z) {=} \sqrt{\frac{2\pi}{R}} \!\!\sum_{n\in\mathbb{Z}+\alpha}\!\!\! \sqrt{\tfrac{\omega_n^+}{2\omega_n}} \!\!\left(\!\! \Psi_ne^{-\frac{i\pi}{4}}e^{\tfrac{i}{2}(\omega_n^+z-\omega_n^-\bar z)} {+} \Psi_n^\dagger e^{\frac{i\pi}{4}}e^{-\tfrac{i}{2}(\omega_n^+z-\omega_n^-\bar z)} \!\!\right).
\end{align}
Again the periodicity condition is now expressed as
\begin{align}
    \Psi(z+R,\bar z+R)=e^{2\pi i\a}\Psi(z,\bar z)\;,\;\bar\Psi(z+R,\bar z+R)=e^{2\pi i\a}\bar\Psi(z,\bar z)\;.
\end{align}
The equations of motion are
\begin{align}
    &\partial\bar\Psi=-\tfrac{1}{2}m\Psi\;,\\
    &\bar\partial\Psi=-\tfrac{1}{2}m\bar\Psi\;,
\end{align}
and these formally come from the Lagrangian density
\begin{align}
    \mathcal{L}=\Psi\bar\partial\Psi-\bar\Psi\partial\bar\Psi+m\Psi\bar\Psi\;.
\end{align}
This is the same Lagrangian density as appears in \cite{Ghoshal:1993tm} but differs by a factor of 2 and the sign of $m$ from that in \cite{Saleur1987} so that here $m>0$ corresponds to the low-temperature phase. As for the Dirac case the Lagrangian density is invariant under a coordinate rotation given by
\be
    (z,\bar z)\rightarrow(e^{i\theta}z,e^{-i\theta}\bar z)\;,
\ee
if the fields transform as
\be
    \Psi(z,\bar z)=e^{\frac{i\theta}{2}}\Psi(e^{i\theta}z,e^{-i\theta}\bar z)\;,\;\bar\Psi(z,\bar z)=e^{-\frac{i\theta}{2}}\bar\Psi(e^{i\theta}z,e^{-i\theta}\bar z)\;.
\ee
Hence bilinear fields of the form $\Psi\partial^{2n-1}\Psi$ and $\bar\Psi\bar\partial^{2n-1}\bar\Psi$ transform under coordinate rotations as
\begin{align}
    &\Psi\partial^{2n-1}\Psi\rightarrow e^{-2n\ii\theta}\Psi\partial^{2n-1}\Psi\;,\\
    &\bar\Psi\bar\partial^{2n-1}\bar\Psi\rightarrow e^{2n\ii\theta}\bar\Psi\bar\partial^{2n-1}\bar\Psi\;.
\end{align}

\section{Uniqueness of tensor currents}\label{Current Uniqueness}
Here we show that, up to total derivatives and powers of the mass~$m$, all bilinear fermion currents take the form of \eqref{Tchoice}. We also show these bilinear currents can be arranged into a conserved, symmetric tensor.

If we only want to construct bilinear currents from the Majorana fields $\bar\Psi$, $\Psi$ and derivatives of these two fields then the most general currents take the form
\begin{align}
    \partial^a\bar\partial^b\Psi\partial^c\bar\partial^d\Psi\;,\;\partial^a\bar\partial^b\bar\Psi\partial^c\bar\partial^d\Psi\;,\;\partial^a\bar\partial^b\bar\Psi\partial^c\bar\partial^d\bar\Psi\;,
\end{align}
for non-negative integers $a,b,c,d$. Up to total derivatives these currents are equivalent to
\begin{align}
    (-1)^{a+b}\Psi\partial^{a+c}\bar\partial^{b+d}\Psi\;,\;(-1)^{a+b}\bar\Psi\partial^{a+c}\bar\partial^{b+d}\Psi\;,\;(-1)^{a+b}\bar\Psi\partial^{a+c}\bar\partial^{b+d}\bar\Psi\;.
\end{align}
We drop the factor $(-1)^{a+b}$ and set $n=a+c$ and $k=b+d$ (so $n$ and $k$ are non-negative integers). The currents are
\begin{align}\label{eq:posscurrents}
    \Psi\partial^n\bar\partial^k\Psi\;,\;\bar\Psi\partial^n\bar\partial^k\Psi\;,\;\bar\Psi\partial^n\bar\partial^k\bar\Psi\;.
\end{align}
Now we use the equations of motion
\be
    \partial\bar\Psi=-\lambda\Psi\;, \quad 
    \bar\partial\Psi=-\lambda\bar\Psi\;,
\ee
where we have set $m=2\lambda$.
First let us assume that $n>k$. Then using the equations of motion gives
\begin{align}
    &\Psi\partial^n\bar\partial^k\Psi=\bfmtb^{2k}\Psi\partial^{n-k}\Psi\;,\\
    &\bar\Psi\partial^n\bar\partial^k\Psi=\bfmtb^{2k}\bar\Psi\partial^{n-k}\Psi\;,\\
    &\bar\Psi\partial^n\bar\partial^k\bar\Psi=-\bfmtb^{2k+1}\bar\Psi\partial^{n-k-1}\Psi\;.\label{eq:bottomcurrent}
\end{align}
Again using the equations of motion, up to total derivatives the currents are
\begin{align}
    &\Psi\partial^n\bar\partial^k\Psi=\bfmtb^{2k}\Psi\partial^{n-k}\Psi\;,\\
    &\bar\Psi\partial^n\bar\partial^k\Psi=\bfmtb^{2k+1}\Psi\partial^{n-k-1}\Psi+\partial(\dots)\;,\\
    &\bar\Psi\partial^n\bar\partial^k\bar\Psi=-\bfmtb^{2k+2}\Psi\partial^{n-k-2}\Psi+\partial(\dots)\;.
\end{align}
Note if $n-k-1=0$ then the current \eqref{eq:bottomcurrent} is
\begin{align}
    -\bfmtb^{2k+1}\bar\Psi\Psi\;.
\end{align}
Alternatively if $k > n$ in \eqref{eq:posscurrents}, using the equations of motion we find that up to total derivatives we have the three currents
\begin{align}
    &\Psi\partial^n\bar\partial^k\Psi=-\bfmtb^{2n+2}\bar\Psi\,\bar\partial^{k-n-2}\bar\Psi+\bar\partial(\dots)\;,\label{eq:topcurrent}\\
    &\bar\Psi\partial^n\bar\partial^k\Psi=-\bfmtb^{2n+1}\bar\Psi\,\bar\partial^{k-n-1}\bar\Psi\;,\\
    &\bar\Psi\partial^n\bar\partial^k\bar\Psi=\bfmtb^{2n}\bar\Psi\,\bar\partial^{k-n}\bar\Psi\;.
\end{align}
Note if $k-n-1=0$ then the current \eqref{eq:topcurrent} is
\begin{align}
    -\bfmtb^{2n+1}\Psi\bar\Psi\;.
\end{align}
Finally for  $n=k$ in \eqref{eq:posscurrents}, using the equations of motion, we see that the only non-vanishing current is
\begin{align}
    \bfmtb^{2k}\bar\Psi\Psi\;.
\end{align}
Hence up to powers of the mass and total derivatives all currents take the form
\be
    \bar\Psi\,\bar\partial^n\bar\Psi \;,\quad \Psi\partial^n\Psi \;,\quad \bar\Psi\Psi\;.
\ee
Additionally, we have
\be
    \bar\Psi\,\bar\partial^{2n}\bar\Psi=\bar\partial \!\left(\sum_{k=1}^n\bar\partial^{k-1}\bar\Psi\bar\partial^{2n-k}\bar\Psi \!\!\right) ,\;  \Psi\partial^{2n}\Psi=\partial \!\left(\sum_{k=1}^n\partial^{k-1}\Psi\partial^{2n-k}\Psi \!\!\right),
\ee
so that for $n$ even the currents are total derivatives. 
Hence the non-trivial currents take the form
\be
    \bar\Psi\,\bar\partial^{2n-1}\bar\Psi \;,\quad \Psi\partial^{2n-1}\Psi \;,\quad \bar\Psi\Psi\;.
\ee
Now these currents can be arranged in a conserved symmetric tensor. We have the conservation equation
\begin{align}
    \partial(\bar\Psi\,\bar\partial^{2n-1}\bar\Psi)+\bar\partial(\lambda\Psi\bar\partial^{2n-2}\bar\Psi)\=0\;,
\end{align}
and up to total derivatives we have
\begin{align}
    \lambda\Psi\bar\partial^{2n-2}\bar\Psi\=\lambda^2\bar\Psi\,\bar\partial^{2n-3}\bar\Psi+\bar\partial(\dots)\;.
\end{align}
Hence up to total derivatives we have the conserved current
\begin{align}
    (\bar J^z,\bar J^{\bar z})=(\bar\Psi\,\bar\partial^{2n-1}\bar\Psi,\lambda^2\bar\Psi\,\bar\partial^{2n-3}\bar\Psi)\;.
\end{align}
Note for $n=1$ then the conserved current becomes
\begin{align}
    (\bar J^z,\bar J^{\bar z})=(\bar\Psi\,\bar\partial\,\bar\Psi,\lambda\Psi\bar\Psi)\;.
\end{align}
Similarly we have
\begin{align}
    &\bar\partial(\Psi\partial^{2n-1}\Psi)+\partial(\lambda\bar\Psi\partial^{2n-2}\Psi)=0\;,\\
    &\lambda\bar\Psi\partial^{2n-2}\Psi=\lambda^2\Psi\partial^{2n-3}\Psi+\partial(\dots)\;,
\end{align}
so up to total derivatives we have the conserved current
\begin{align}
    (J^z,J^{\bar z})=(\lambda^2\Psi\partial^{2n-3}\Psi,\Psi\partial^{2n-1}\Psi)\;.
\end{align}
and again if $n=1$ the current is
\begin{align}
    (J^z,J^{\bar z})=(\lambda\bar\Psi\Psi,\Psi\partial\Psi)\;.
\end{align}
These calculations show that the following symmetric rank-$2n$ tensor
\be
T_{(2n,k)}\= \begin{cases} \frac{i}{2}\bfmtb^{2n+k}\bar\Psi\,{\bar\partial}^{-k-1}\bar\Psi\;,\qquad & k\=-2n,-2n+2,\dots,-2\;,\\
\frac{i}{2}\bfmtb^{2n-1}\Psi\bar\Psi\;,\qquad & k\=0\;,\\
-\frac{i}{2}\bfmtb^{2n-k}\Psi\partial^{k-1}\Psi & k\=2,4,\dots,2n \,,
\end{cases}
\ee
is conserved, i.e.
\be
    \partial T_{(2n,k)} + \bar\partial T_{(2n,k+2)} \= 0\;.
\ee

\section{Regularisation of constants}\label{regularisation}

In this appendix we calculate the constant terms $E_0$ and $C$ that appear in the Hamiltonian \eqref{eq:HandP} and one point functions \eqref{eq:ToneptD} respectively. This is done via a zeta-function style regularisation process and the values we obtain respect the modular properties of the partition and one point functions.

\subsection{Regularising $E_0$}
The Hamiltonian \eqref{eq:HandP} has a constant term formally given by the divergent series
\be
    -\frac{1}{2}\sum_{k\in\Zint}\sqrt{m^2 + \left(\tfrac{2\pi(k+\a)}{R}\right)^2}\;.
\ee
In order to make sense of this expression, we consider 
the following sum which converges for Re$(\nu)>0$ 
\be
    E_0(\nu)\=-\frac{1}{2}\sum_{k\in\Zint} \left(m^2 + \left(\tfrac{2\pi(k+\a)}{R}\right)^2 \right)^{-\nu-\tfrac{1}{2}}\;.
\ee
This series is periodic in $\a$ with period 1, and so it has a Fourier expansion
\be
    E_0(\nu)\=\sum_{\ell\in\Z}a_\ell \, e^{2\pi \ii \ell \a}\,.
\ee
The coefficients $a_\ell$ are given by
\be \label{aellcalc}
\begin{split}
    a_\ell &\=-\frac{1}{2}\int_0^1 d\a\, e^{-2\pi \ii \ell \a}\sum_{k\in\Z} \left(m^2 + \left(\tfrac{2\pi(k+\a)}{R}\right)^2\right)^{-\nu-\tfrac{1}{2}} \\
    & \=-\int_0^\infty d\a\, \lb m^2 + \lb \tfrac{2\pi\a}{R} \rb^2 \rb^{-\nu-\tfrac{1}{2}}\cos(2\pi \ell \a)\;,
    \end{split}
\ee
where, in going to the second line, we have used
\be \label{unfolding}
    \int_0^1 d\a\,\sum_{k\in\Z}f(k+\a)=\int_{-\infty}^\infty d\a\,f(\a)
\ee
for any integrable function~$f$, 
combined with the fact the integrand is even in $\a$. 
The integral in~\eqref{aellcalc} can be evaluated,
for Re$(m)>0$ and Re$(\nu)>-\tfrac{1}{2}$, to be ((7) on page 11 of \cite{Manuscript1955TablesOI})
\be\begin{split}
   & \lb \frac{R}{2\pi}\rb^{2\nu+1} \int_0^\infty d\a\,\lb \a^2+\lb \tfrac{mR}{2\pi}\rb^2 \rb^{-\nu-\frac{1}{2}}\cos(2\pi\ell \a) =\\
&\frac{R}{2\sqrt{\pi}\Gamma(\nu+\frac{1}{2})}\lb \frac{R|\ell|}{2m}\rb^\nu K_\nu(mR|\ell|)\;,
\end{split}
\ee
where $K_\nu(x)$ is the modified Bessel function of the second kind. We thus obtain, for~$\ell \neq 0$, 
\be
    a_\ell \= -\frac{R}{2\sqrt{\pi}\Gamma(\nu+\frac{1}{2})}\lb \frac{R|\ell|}{2m}\rb^\nu K_\nu(mR|\ell|)\;.
\ee
The coefficient $a_0$ is defined as a limit~$\ell \to 0$ of the above expression considering~$\ell$ as a continuous real variable:
\be
    a_0 \= -\frac{R\,\Gamma(\nu)}{4\sqrt{\pi}\,m^{2\nu}\,\Gamma(\nu+\frac{1}{2})}\;.
\ee
Thus we obtain the following expression,
\be\label{eq:E0continued}
    E_0(\nu) = -\frac{R\,\Gamma(\nu)}{4\sqrt{\pi}\,m^{2\nu}\,\Gamma(\nu{+}\frac{1}{2})} - \frac{R^{\nu+1}}{(2m)^\nu \sqrt{\pi}\,\Gamma(\nu{+}\frac{1}{2})}\sum_{\ell = 1}^\infty\! \ell^\nu K_\nu(2\pi m \ell)\cos(2\pi \ell \a).
\ee
The series in the second term is convergent for all $\nu \in \C$. 
The function $E_0(\nu)$ as given in~\eqref{eq:E0continued} is therefore a meromorphic function in $\nu \in \C$ with simple poles at $\nu=0,-1,-2,\dots$.  
If we subtract the $a_0$ term and analytically continue to $\nu=-1$ we find
\be
    E_0 = \lim_{\nu\to -1}\left(E_0(\nu)+\frac{R\,\Gamma(\nu)}{4\sqrt{\pi}\,m^{2\nu}\,\Gamma(\nu{+}\frac{1}{2})}\right) 
    = \frac{m}{\pi}\sum_{\ell = 1}^\infty\frac{1}{\ell} K_1(mR \ell) \cos(2\pi \ell \a)\;.
\ee
where we have used $K_{-1}(x)=K_1(x)$ and $\Gamma(-\tfrac{1}{2})=-2\sqrt{\pi}$. 
This is the regularised value of the zero-point energy~$E_0$ that we use, which as mentioned below \eqref{eq:E0}, ensures that the partition function is modular invariant.

\subsection{Regularising $C$}

Similarly, when calculating the one-point functions \eqref{eq:ToneptD} for the fermions we have constant terms formally given by the divergent series 
\be
\sum_{k\in\Z}\frac{\lb\sqrt{m^2 {+} \lb\tfrac{2\pi(k+\a)}{R}\rb^2} {+} \frac{2\pi(k+\a)}{R}\rb^{2n} {+} \lb\sqrt{m^2 {+} \lb\tfrac{2\pi(k+\a)}{R}\rb^2} {-}  \frac{2\pi(k+\a)}{R}\rb^{2n}}{2\sqrt{m^2 {+} \lb\tfrac{2\pi(k+\a)}{R}\rb^2}}\;,
\ee
for Dirac fermions. In the Majorana case we have $\a=0,\tfrac{1}{2}$ and the sum can be written as
\be
\sum_{k\in\Z}\frac{\lb\sqrt{m^2+\lb\tfrac{2\pi(k+\a)}{R}\rb^2} \pm \frac{2\pi(k+\a)}{R}\rb^{2n}}{\sqrt{m^2+\lb\tfrac{2\pi(k+\a)}{R}\rb^2}}\;.
\ee
As above, we regulate this series by the parameters~$\nu$ and~$\mu$,
\be\begin{split}
   & C(\nu,\mu) = \\
&\sum_{k\in\Z}\! \frac{ \!\lb\!\! \sqrt{m^2{+} \!\lb\! \tfrac{2\pi(k+\a)}{R} \!\rb^2}{+}\frac{2\pi(k+\a)}{R} \!\!\rb^\nu \!\!{+} \!\lb\!\! \sqrt{m^2{+}\!\lb\!\tfrac{2\pi(k+\a)}{R}\!\rb^2}{-} \frac{2\pi(k+\a)}{R}\!\!\rb^\nu}{2\sqrt{m^2+\lb\!\tfrac{2\pi(k+\a)}{R}\!\rb^2}} e^{2\pi \ii \mu(k+\a)},
\end{split}
\ee
which converges for $|$Re$(\nu)|<1$ and $\mu\in(0,1)$. 
Note that if $\mu = 0,1$ then $C(\nu,\mu)$ does not converge for any $\nu$. 
This function is periodic in $\a$ with period 1 and so has a Fourier expansion, 
\be
    C(\nu,\mu) \= \sum_{\ell\in\Z} a_\ell e^{2\pi \ii \ell \a}\;.
\ee
The coefficients $a_\ell$ are given by
\be
\begin{split}\label{eq:al}
    &a_\ell = \int_0^1\! d\a \sum_\pm \sum_{k\in\Z} \frac{ \lb \sqrt{m^2{+} \lb \tfrac{2\pi(k+\a)}{R} \!\rb^2} \pm \frac{2\pi(k+\a)}{R} \rb^\nu}{2\sqrt{m^2+\lb \tfrac{2\pi(k+\a)}{R} \rb^2}} e^{2\pi \i (\mu - \ell) (k+\a)}\;,\\
    &= \int_0^\infty\!\! d\a \frac{\lb\!\sqrt{m^2 {+} \lb\tfrac{2\pi\a}{R}\rb^2} {+} \frac{2\pi\a}{R}\!\rb^\nu {+} \lb\!\sqrt{m^2 {+} \lb\tfrac{2\pi\a}{R}\rb^2} {-} \frac{2\pi\a}{R}\!\rb^\nu}{\sqrt{m^2 + \lb\tfrac{2\pi\a}{R}\rb^2}} \cos(2\pi (\ell {-} \mu) \a)\;,
\end{split}
\ee
where we have unfolded the sum as in~\eqref{unfolding}, 
and used the fact the integrand is even in $\a$. We use the integral identity ((22) on page 13 of \cite{Manuscript1955TablesOI})
\be\begin{split}
    &\lb\frac{2\pi}{R}\rb^{\nu-1} \!\int_0^\infty \!d\a \frac{\lb\!\! \sqrt{\lb\tfrac{mR}{2\pi}\rb^2 {+} \a^2} + \!\a \!\!\rb^\nu \!{+} \lb\!\! \sqrt{\lb\tfrac{mR}{2\pi}\rb^2 {+} \a^2} - \!\a \!\!\rb^\nu}{\sqrt{\lb\tfrac{mR}{2\pi}\rb^2 + \a^2}} \cos(2\pi (\ell {-} \mu) \a)\;,\\
    &\= \frac{m^\nu R}{\pi} \cos\left(\frac{\nu \pi}{2}\right) K_{\nu}(mR |\ell-\mu|)
    \end{split}
\ee
which is valid for Re$(m)>0$ and $|$Re$(\nu)|<1$ (since $\mu\in(0,1)$ we never have $\ell = \mu$). 
We find
\be
    C(\nu,\mu) \= \frac{m^\nu R}{\pi} \cos\left(\frac{\nu\pi}{2}\right) \sum_{\ell \in \Z} K_\nu(mR |\ell-\mu|) e^{2\pi \ii \ell \a}\;.
\ee
which is a convergent sum for all $\nu \in \IC$  when $\mu\in(0,1)$. 

Now we want to take the limit~$\mu\to0$. 
In this limit, the $a_\ell$ are all well-defined except for $a_0$ which diverges.
As before, we define~$C(\nu)$ by subtracting the $a_0$ term,
\be\begin{split}
    C(\nu) &\= \lim_{\mu \to 0}\left(C(\nu,\mu)- \frac{m^\nu R}{\pi} \cos\left(\frac{\nu\pi}{2}\right) K_\nu(2\pi m|\mu|)\right) \\
&\= \frac{2m^\nu R}{\pi} \cos\left(\frac{\nu\pi}{2}\right) \sum_{\ell = 1}^\infty K_\nu(mR \ell) \cos(2\pi \ell \a)\;.
\end{split}
\ee
Finally, we analytically continue to $\nu = 2n$,
\be
    C \= C(2n) \= \frac{2(-1)^n m^{2n}R}{\pi} \sum_{\ell = 1}^\infty K_{2n}(mR \ell) \cos(2\pi \ell \a)\;.
\ee
As before, this value of the regularised constant term ensures that the one-point functions are modular forms.

\section{Sum over Bessel functions}\label{Bessel sum}
In this appendix we derive an expression for the one point functions 
as a lattice sum of Bessel functions, which makes the modular properties obvious. We follow the same procedure as was used in \cite{Berg:2019jhh}.

We start with the expression given in~\eqref{eq:onepoint},
\begin{equation}
\sum_{k\in\mathbb{Z}}\frac{(\sqrt{(2\pi(k {+} \alpha))^2 {+} m^2}+2\pi(k {+} \alpha))^{2n}}{\sqrt{(2\pi(k+\alpha))^2+m^2}}\text{Li}_0 \!\left(\! e^{-\tau_2\sqrt{(2\pi(k+\alpha))^2+m^2}+2\pi i\tau_1(k+\alpha)+2\pi i\beta}\!\right),
\end{equation}
where $\text{Li}_0(x)=\tfrac{x}{1-x}$.
This function is clearly periodic in $\alpha$ and $\beta$ with period 1 so we can write it as a Fourier expansion of the form
\begin{equation}
\sum_{(r,\ell)\in\mathbb{Z}^2}a_{r,\ell} \, \rme^{2\pi i(\alpha \ell+\beta r)}\;.
\end{equation}
To find the coefficients $a_{r,\ell}$ we need to compute the integral
\begin{equation}
\int_0^1d\alpha d\beta e^{-2\pi i(\ell\alpha+r\beta)} \sum_{k\in\mathbb{Z}} \frac{(\omega_{k+\a}^+)^{2n}}{\omega_{k+\a}}\text{Li}_0\!\left(\!e^{-\tau_2\omega_{k+\a}+2\pi i\tau_1(k+\alpha)+2\pi i\beta}\right)\,,
\end{equation}
where
\be
	\omega_k = \sqrt{(2\pi k)^2 + m^2} \;,\quad \omega_k^+ = \omega + 2\pi k\;.
\ee
We expand the Li$_0$ as a power series and rearrange to get
\begin{equation}
\sum_{k\in\mathbb{Z}}\sum_{j=1}^\infty\int_0^1 \!\!d\beta e^{2\pi i(j-r)\beta}\int_0^1 \!\! d\alpha\frac{(\omega_{k+\a}^+)^{2n}}{\omega_{k+\a}}e^{2\pi i(\tau_1(k+\alpha)j-\ell\alpha)-\tau_2 j \omega_{k+\a}}\;.
\end{equation}
The $\beta$ integral gives us a Kronecker delta
\begin{equation}
\sum_{k\in\mathbb{Z}}\sum_{j=1}^\infty\delta_{j,r}\int_0^1 d\alpha \frac{(\omega_{k+\a}^+)^{2n}}{\omega_{k+\a}}e^{2\pi i(\tau_1(k+\alpha)j-\ell\alpha)-\tau_2 j \omega_{k+\a}}\;.
\end{equation}
Note that since $j>0$, $a_{r,l}$ vanishes for $r\leq0$. For the rest of the calculation we assume $r>0$. The Kronecker delta kills the $j$ sum
\begin{equation}
\sum_{k\in\mathbb{Z}}\int_0^1 \!\!d\alpha\frac{(\sqrt{(2\pi(k{+}\alpha))^2 {+} m^2} +\! 2\pi(k {+} \alpha)\!)^{2n}}{\sqrt{(2\pi(k {+} \alpha))^2+m^2}}e^{2\pi i(\tau_1(k+\alpha)r-\ell\alpha)-\tau_2r\sqrt{(2\pi(k+\alpha))^2+m^2}}\;.
\end{equation}
We can replace the sum over integers with an integral over the reals 
using the identity
\begin{equation}
\sum_{k\in\mathbb{Z}}\int_0^1d\alpha f(k+\alpha)=\int_{-\infty}^\infty d\alpha f(\alpha)\;.
\end{equation}
Using this we find
\be\begin{split}
& \int_{-\infty}^\infty d\alpha \frac{(\sqrt{(2\pi\alpha)^2+m^2}+2\pi\alpha)^{2n}}{\sqrt{(2\pi\alpha)^2+m^2}}e^{2\pi i(\tau_1r - \ell)\alpha-\tau_2r\sqrt{(2\pi\alpha)^2+m^2}}\;,\\
&(2\pi)^{2n-1}\!\! \int_{-\infty}^\infty \!d\alpha \frac{(\sqrt{\a^2 {+} (\frac{m}{2\pi})^2} {+} \alpha)^{2n}}{\sqrt{\a^2 {+} (\frac{m}{2\pi})^2}}e^{-2\pi i \ell\alpha+\pi i\tau r(\alpha+\sqrt{\alpha^2+(\frac{m}{2\pi})^2})+\pi i\bar\tau r(\alpha-\sqrt{\alpha^2+(\frac{m}{2\pi})^2})}\;,
\end{split}
\ee
where $\t = \t_1 + \ii \t_2$. We can exchange the $(\sqrt{\alpha^2+(\frac{m}{2\pi})^2}+\alpha)^{2n}$ term for $\tau$ derivatives
\begin{align}
\frac{(-1)^n2^{2n-1}}{\pi\, r^{2n}}\frac{\partial^{2n}}{\partial\tau^{2n}}\int_{-\infty}^\infty d\alpha \frac{1}{\sqrt{\alpha^2+(\frac{m}{2\pi})^2}}e^{2\pi i(\tau_1r- \ell)\alpha-2\pi\tau_2r\sqrt{\alpha^2+(\frac{m}{2\pi})^2}}\;.\label{eq:tau derivatives}
\end{align}
We have the following integral identity for modified Bessel functions of the second kind ((27) on Page 17 of \cite{Manuscript1955TablesOI}),
\begin{equation}
\int_0^\infty \!\!\!d\alpha \frac{1}{\sqrt{\alpha^2{+}(\frac{m}{2\pi})^2}}\cos(2\pi \a (\t_1 r {-} \ell)\!)e^{-2\pi \t_2 r \sqrt{\alpha^2+(\frac{m}{2\pi})^2}} {=} K_0(\! m |\t r {-} \ell|)\;,
\end{equation}
for Re$(m)>0$ and Re$(\t_2 r)>0$. This gives us
\begin{align}
a_{r,\ell} \= \frac{(-1)^n2^{2n}}{\pi\,r^{2n}}\frac{\partial^{2n}}{\partial\tau^{2n}}K_0(2\pi m|r\tau - \ell|)\;.
\end{align}
Using the following integral representation ((29) on page 146 of \cite{Manuscript1955TablesOI})
\begin{align}
K_{2n}(x)=\int_0^\infty ds\frac{x^{2n}}{(2s)^{2n+1}}e^{-s-\frac{x^2}{4s}}\;,\label{eq:bessel int}
\end{align}
for Re$(x^2)>0$, one can show
\begin{align}
\frac{\partial^{2n}}{\partial\tau^{2n}}K_0(m|r\tau - \ell|)=\lb\frac{mr}{2}\rb^{2n}\left(\frac{\bar\tau r - \ell}{\tau r - \ell}\right)^nK_{2n}(m|r\tau - \ell|)\;,
\end{align}
which gives
\begin{align}
a_{r,\ell} \= \begin{cases} 0\;, & r\leq 0\;,\\
    \frac{(-1)^nm^{2n}}{\pi}\left(\frac{\bar\tau r - \ell}{\tau r - \ell}\right)^nK_{2n}(m|r\tau - \ell|)\;, & r>0\;.
    \end{cases}
\end{align}
Thus we obtain the Fourier expansion
\be\begin{split}
&\sum_{k\in\mathbb{Z}}\!\frac{(\sqrt{(2\pi(k{+}\alpha))^2 {+} m^2} +\! 2\pi(k {+} \alpha)\!)^{2n}}{\sqrt{(2\pi(k+\alpha))^2+m^2}}\text{Li}_0 \!\left(\! e^{-\t_2\sqrt{(2\pi(k+\alpha))^2+m^2}+2\pi i\tau_1(k+\alpha)+2\pi i\beta} \!\right)\\
& \= \frac{(-1)^nm^{2n}}{\pi}\sum_{\begin{smallmatrix} \ell \in\mathbb{Z}\\r>0\end{smallmatrix}}\left(\frac{\bar\tau r - \ell}{\tau r - \ell}\right)^nK_{2n}(m|r\tau - \ell|)e^{2\pi \ii(\alpha \ell + \beta r)}\,.
\end{split}\label{eq:halflattice}
\ee
Restricting to $\alpha, \beta \in \{0,\tfrac{1}{2}\}$ allows us to write the Fourier transform as a lattice sum plus a constant term
\be\begin{split}
&\sum_{k\in\mathbb{Z}}\frac{(\sqrt{(2\pi(k {+} \alpha))^2 {+} m^2} +\! 2\pi(k {+} \alpha))^{2n}}{\sqrt{(2\pi(k {+} \alpha))^2 {+} m^2}}\text{Li}_0 \!\left(\! e^{-\tau_2\sqrt{(2\pi(k+\alpha))^2+m^2}+2\pi i\tau_1(k+\alpha)+2\pi i\beta} \!\right)\\
&= \frac{(-1)^nm^{2n}}{2\pi} \!\!\left( \sideset{}'\sum_{(r,\ell)\in\mathbb{Z}^2} \!\!\!\left(\! \frac{\bar\tau r {-} \ell}{\tau r {-} \ell} \!\right)^n \!\!K_{2n}(m|r\tau {-} \ell|)e^{2\pi i(\alpha \ell+\beta r)} {-} 2\sum_{\ell = 1}^\infty \!K_{2n}(m \ell)\cos(2\pi \alpha \ell) \!\!\right),
\end{split}
\ee
where we exclude the origin from the lattice sum. If, instead, we start with
\begin{equation}\label{eq:minusk}
\sum_{k\in\mathbb{Z}}\frac{(\sqrt{(2\pi(k {+} \alpha))^2 {+} m^2} {-} 2\pi(k {+} \alpha))^{2n}}{\sqrt{(2\pi(k+\alpha))^2+m^2}}\text{Li}_0 \!\left(\! e^{-\tau_2\sqrt{(2\pi(k+\alpha))^2+m^2}+2\pi i\tau_1(k+\alpha)+2\pi i\beta} \!\right),
\end{equation}
then all the above calculations go through, except that in \eqref{eq:tau derivatives} we replace the $\tau$-derivatives 
with $\bar\tau$-derivatives, so that the final result is
\be\begin{split}
&\sum_{k\in\mathbb{Z}}\frac{(\sqrt{(2\pi(k {+} \alpha))^2 {+} m^2} -\! 2\pi(k {+} \alpha))^{2n}}{\sqrt{(2\pi(k {+} \alpha))^2 {+} m^2}}\text{Li}_0 \!\left(\! e^{-\tau_2\sqrt{(2\pi(k+\alpha))^2+m^2}+2\pi i\tau_1(k+\alpha)+2\pi i\beta} \!\right)\\
&= \frac{(-1)^nm^{2n}}{2\pi} \!\!\left( \sideset{}'\sum_{(r,\ell)\in\mathbb{Z}^2} \!\!\!\left(\! \frac{\tau r {-} \ell}{\bar\tau r {-} \ell} \!\right)^n \!\!K_{2n}(m|r\tau {-} \ell|)e^{2\pi i(\alpha \ell+\beta r)} {-} 2\sum_{\ell = 1}^\infty \!K_{2n}(m \ell)\cos(2\pi \alpha \ell) \!\!\right),
\end{split}
\ee
Note that if we symmetrise \eqref{eq:halflattice} in $\a$ and $\b$ then for $\a,\b \in \R$ we have the lattice sum
\be\begin{split}
&\sum_{k\in\mathbb{Z}}\frac{(\sqrt{(2\pi(k {+} \alpha))^2 {+} m^2} +\! 2\pi(k {+} \alpha))^{2n}}{\sqrt{(2\pi(k {+} \alpha))^2 {+} m^2}}\text{Li}_0 \!\left(\! e^{-\tau_2\sqrt{(2\pi(k+\alpha))^2+m^2}+2\pi i\tau_1(k+\alpha)+2\pi i\beta} \!\right)\\
&= \frac{(-1)^nm^{2n}}{\pi} \!\!\left( \sideset{}'\sum_{(r,\ell)\in\mathbb{Z}^2} \!\!\!\left(\! \frac{\bar\tau r {-} \ell}{\tau r {-} \ell} \!\right)^n \!\!K_{2n}(m|r\tau {-} \ell|)e^{2\pi i(\alpha \ell+\beta r)} {-} 2\sum_{\ell = 1}^\infty \!K_{2n}(m \ell)\cos(2\pi \alpha \ell) \!\!\right),
\end{split}
\ee
We can similarly symmetrise \eqref{eq:minusk} in $\a$ and $\b$ to obtain a lattice sum for $\a,\b \in \R$.
These symmetrised expressions are precisely the one-point functions of
the Dirac fermions as discussed in Section~\ref{sec:Diracsymm}. 

\bibliographystyle{imsart-number.bst} 
\bibliography{FreeFermionsRefs}

\end{document}